\documentclass[aps,pra,reprint,superscriptaddress,nofootinbib]{revtex4-2}
\usepackage[OT2,T1]{fontenc}
\usepackage[utf8]{inputenc}
\usepackage{amsmath}
\usepackage{amssymb}
\usepackage{graphicx}
\usepackage{hyperref}
\usepackage{xcolor}
\usepackage{physics}

\DeclareSymbolFont{cyrletters}{OT2}{wncyr}{m}{n}
\DeclareMathSymbol{\Sha}{\mathalpha}{cyrletters}{"58}

\begin{document}

\title{Statistical Mechanics of Spike Trains: Self-Consistent Activity Ensembles and Phase Transitions}

\author{Jan Pyrzowski }
\affiliation{Université Paris Cité, Institute of Psychiatry and Neuroscience of Paris (IPNP), INSERM U1266, Neuronal Signaling in 
Epilepsy and Glioma, Paris, France}
\affiliation{Hôpital Fondation Adolphe de Rothschild, service de Neurologie, Paris, France}
\affiliation{Department of Emergency Medicine, Medical University of Gdańsk, Poland}
\author{Laura Messio}
\affiliation{Sorbonne Université, CNRS, Laboratoire de Physique Théorique de la Matière Condensée, LPTMC, F-75005 Paris, France}
\affiliation{ Institut Universitaire de France (IUF), F-75005 Paris, France}
\author{Guillaume Marrelec}
\affiliation{LIB, Sorbonne Université, INSERM, CNRS, 75006 Paris, France}
\author{Marcin Markiewicz }
\affiliation{Institute of Theoretical and Applied Informatics, Polish Academy of
Sciences, ul. Ba{\l}tycka 5, 44-100 Gliwice, Poland} \affiliation{International
Centre for Theory of Quantum Technologies (ICTQT), University of Gdansk, 80-308
Gdansk, Poland}
\date{\today}

\begin{abstract}
    We introduce a statistical-mechanical framework for spiking neuronal systems in which entire spike trains, defined as spatiotemporal configurations of discrete events, are treated as microstates of the system. Instead of modeling explicit network dynamics, we construct an ensemble over complete spike histories and assign to each configuration an effective energy functional that quantifies its consistency with intrinsic synaptic interactions.

This formulation leads to a Boltzmann-like distribution over spike configurations, where an effective temperature captures the influence of external inputs acting as a stochastic drive. Macroscopic states are described in terms of coarse-grained observables, such as population firing-rate densities, and arise from a competition between energetic contributions favoring interaction-consistent spike patterns and entropic contributions favoring spatio-temporal disorder.

Within this framework, we primarily characterize the transition between
oscillatory and non-oscillatory activity. The non-oscillatory regime
contains both broadly distributed, approximately uniform states
and temporally localized collapsed states, whereas the
oscillatory regime is characterized by repeated activity peaks. These regimes are diagnosed
using order parameters derived from the population spike-rate density,
including its entropy, temporal localization measures, and
autocorrelation-based indicators.

To complement the analytical treatment, we perform Monte Carlo simulations that sample the ensemble of admissible spike configurations using a Metropolis–Hastings scheme defined directly on spike trains. The simulations reveal the emergence of the predicted macroscopic regimes and provide a direct sampling-based probe of the free-energy landscape.

\end{abstract}

\maketitle

\section{Introduction}

Spiking neuronal systems are usually described by specifying a
dynamical rule for the evolution of neuronal states and then studying the
spike (action potential) trains generated by this rule. In such approaches, spikes appear as
outputs of an underlying membrane or population dynamics. Here we take a
complementary point of view. We treat complete spike trains themselves as
the microscopic states of the system and define a statistical
ensemble directly on the space of such configurations. This
trajectory-based perspective is inspired by probabilistic formulations of
dynamical systems in which deterministic evolution laws are represented
through ensembles of evolution paths or histories \cite{Constantin08}.

In this formulation, a microstate is a spatiotemporal pattern of
discrete spike events indexed by neuron identity and time over a finite
observation window. The network's internal structure is encoded in an intrinsic response operator, which maps a spike configuration to the endogenous activation field generated through internal membrane-potential dynamics, synaptic weights, transmission delays, and response kernels.
 An effective
energy functional then measures the consistency between the spikes
present in the configuration and the activation field generated by the
same configuration. Spike configurations are assigned lower energy when their events coincide
with strong internally generated activation, while configurations that
are poorly supported by the network's intrinsic interactions are
energetically penalized.

This construction leads to a Boltzmann-like ensemble over complete spike
histories. The inverse temperature controls the relative importance of
endogenous network consistency and residual fluctuations attributed to
external drive, channel noise, unobserved inputs, and other influences not
represented explicitly by the network model. In the low-temperature regime the
ensemble concentrates on configurations that are self-consistent with the
intrinsic interactions, while in the high-temperature regime admissible
spike configurations become increasingly dominated by entropy. Although the underlying neuronal dynamics is open and driven, the present
construction uses an equilibrium-like Boltzmann ensemble over complete
spike histories. In this sense, it connects spiking network theory with
the language of trajectory ensembles, maximum-entropy statistical
mechanics, and disordered interacting systems \cite{Seifert12,Aguilera21,Aguilera23}.

Macroscopic activity states are obtained by coarse-graining the
trajectory ensemble. In particular, we use the population spike-rate
density as a central observable and study its entropy, spectral content,
and localization. The competition between the energy favouring
interaction-consistent spike patterns and the entropy favouring temporal
disorder gives rise to distinct macroscopic regimes. In the excitatory
delayed networks analysed below these include temporally uniform states,
delay-induced oscillatory states, and temporally localized
discharge-like states. These regimes are related, at the phenomenological
level, to asynchronous, rhythmic, polychronous, and synchronized
population activity observed in spiking network models \cite{Izhikevich10,Izhikevich06,Brunel00,Vreeswijk96}.

Our approach differs from traditional population-level mean-field
descriptions, which usually derive deterministic or low-dimensional
evolution equations for macroscopic observables such as firing rates or
population voltages \cite{Montbri015,diVolo19,Chen22,Tahvili24}.
Here the ensemble is defined first at the level of complete spike
histories, and mean-field theory is introduced only as a subsequent
coarse-graining. This distinction is important: the microscopic object is
not an instantaneous population dynamical state but a full trajectory, while the
population-rate density is an observable extracted from the ensemble or a
macrostate used in an approximate theory.

The paper is organized as follows. In Sec.~\ref{sec:ensemble} we introduce the general
spike-train ensemble, define the energy functional, and describe two
realizations of the intrinsic response: a biologically motivated
integrate-and-fire construction and a reduced discrete transfer-operator
realization used in the Monte Carlo simulations. In Sec.~\ref{sec:meanfield} we derive a
mean-field reduction in terms of population-rate macrostates. In
Sec.~\ref{sec:phases} we analyse the corresponding mean-field phases of excitatory
delayed networks. In Sec.~\ref{sec:monte_carlo} we sample the underlying finite-size
microscopic ensemble using a Metropolis--Hastings procedure on spike
trains and compare the sampled regimes with the mean-field picture.
In Sec.~\ref{sec:conn} we outline how the same formalism can be inverted for
connectivity inference. Sec.~\ref{sec:disc} contains the discussion and conclusions.

The central aim of this work is not to replace detailed dynamical models of spiking neurons, but to provide a statistical-mechanical description of spike-train ensembles complementary to them. The resulting framework places neural spike patterns in the language of microstates, energies, entropies, free energies, and phase transitions, while retaining the explicit temporal structure of complete spike histories. From this perspective, experimentally observed spike trains can be interpreted as samples from an underlying ensemble shaped by both intrinsic connectivity and external inputs. More generally, by placing spike trains within the broader class of trajectory-based statistical systems, the approach establishes a conceptual bridge between theoretical neuroscience and statistical physics, and suggests a route toward free-energy-based inference, model comparison, and data analysis for experimentally measurable neuronal activity \cite{Andrews24, Giraud25}.

\section{Spike-Train Ensemble: General Framework and Single-Neuron Realizations}
\label{sec:ensemble}

\subsection{Network model}
\label{sec:graph}

We consider a population of $N$ neurons whose connectivity is encoded
by a directed, weighted graph
$\mathcal{G} = (\mathcal{V}, \mathcal{E}, W, \Delta)$, where
$\mathcal{V} = \{1, \dots, N\}$ is the set of neurons and
$\mathcal{E} \subseteq \mathcal{V} \times \mathcal{V}$ the set of
directed synaptic connections. Each edge $(i,j) \in \mathcal{E}$
carries a synaptic weight $W_{ij} \in \mathbb{R}$ (positive for
excitation, negative for inhibition) and a non-negative transmission
delay $\Delta_{ij} \geq 0$. 
In addition to this connectivity structure, the graph is equipped with a single-neuron response rule specifying how incoming synaptic events are converted into an internal activation field and, in more detailed realizations, into membrane-potential dynamics. This rule may range from simple linear response kernels to explicit spiking-neuron models such as integrate-and-fire or more elaborate hybrid dynamical systems \cite{Izhikevich06book}.
We adopt the convention $W_{ij}=0$
whenever $(i,j) \notin \mathcal{E}$ and write
$\mathrm{pre}(j) = \{ i : (i,j) \in \mathcal{E} \}$ for the
presynaptic set of neuron $j$. For any single realization of the
model the graph $\mathcal{G}$ is treated as a quenched structure:
the spike-train ensemble defined in
Sec.~\ref{sec:abstract} is conditioned on $\mathcal{G}$.

The general framework accommodates an arbitrary directed graph and
arbitrary real-valued weights. Two specialisations of this structure
recur throughout the paper, and we introduce them here for later
reference.

\paragraph{Random-graph ensemble.}
In both the analytical mean-field treatment (Secs.~\ref{sec:meanfield}--\ref{sec:phases}) and the
Monte Carlo simulations (Sec.~\ref{sec:monte_carlo}), the graph itself is drawn from a
sparse random-graph ensemble \cite{Erdos60}: each ordered pair $(i,j)$ with $i \neq j$
is independently included in $\mathcal{E}$ with connection probability
$p_{\mathrm{conn}}$, with self-loops excluded. A fresh realization of \(\mathcal G\) is sampled for each Monte Carlo run. In the
numerical phase map below, each point of the parameter grid is represented
by one such quenched graph-and-delay realization and one finite Markov
chain. Thus the plotted observables should be interpreted as
single-realization finite-size samples rather than disorder averages at
fixed parameter values. Disorder-averaged phase diagrams could be
obtained by repeating the simulations over many graph realizations for
each parameter point, but this is not done in the present numerical
study.
\paragraph{Weight and delay distributions.}
Synaptic weights and delays are assigned from prescribed distributions
\(P_W\) and \(P_\Delta\), and are taken to be independent unless stated
otherwise. The analytical mean-field treatment in Secs.~\ref{sec:meanfield}--\ref{sec:phases} and the
Monte Carlo simulations in Sec.~\ref{sec:monte_carlo} focus on homogeneous excitatory
networks, for which we write simply \(P_W\), \(P_\Delta\), and moments
such as \(\langle W\rangle\). More general heterogeneous or mixed-sign
synaptic populations can be incorporated by introducing class-specific
weight and delay distributions, but they are not analysed in the present
work.

The
simplest non-trivial \emph{numerical} setting, adopted in our
Monte Carlo simulations, is the homogeneous excitatory network with
binary unit weights, $P_W(W) = \delta(W - 1)$, so that the quenched
structural disorder enters only through the random adjacency and the
delay assignment. The delay distribution $P_\Delta$ is taken from a
maximum-entropy family on $\mathbb{R}_{\geq 0}$ characterized by its
mean $\mu_\Delta$ and standard deviation $\sigma_\Delta$, which
interpolates between a Gaussian ($\sigma_\Delta \ll \mu_\Delta$) and
an exponential ($\sigma_\Delta \to \mu_\Delta$) limit; both regimes
play distinct roles in the analysis (Secs.~\ref{sec:meanfield}--\ref{sec:phases}). The motivation
for the binary-weight choice is discussed alongside the methodology
in Sec.~\ref{sec:monte_carlo}.

\subsection{Microstates and the abstract ensemble}
\label{sec:abstract}
In the present framework, a spike is not modeled as a detailed voltage waveform or as the full biophysical process of action-potential generation. It is treated as a discrete event assigned to a neuron and to a time. Thus, a spike records the fact that neuron \(i\) emitted an action potential at time \(t\), while the underlying waveform, ionic currents, and subthreshold membrane-voltage dynamics are represented only implicitly through the intrinsic response operator introduced below. This event-based viewpoint is therefore well suited to the trajectory
ensemble: a microstate records which neurons spike and when, while the
underlying voltage traces are represented only implicitly. It also reflects the form in which network activity is often accessed in modern 
electrophysiological experiments, where
network activity is commonly analysed in terms of detected spike events
 and population activity patterns
\cite{Andrews24}.

The fundamental observable of our framework is the population spike
train over a finite observation window $[0,T]$:
\begin{align}
s & \equiv \{ s_i(t) : i = 1, \dots, N \},
\label{eq:microstate} \\
s_i(t) & = \sum_{\substack{t' \in \mathrm{sp}(i) \\ 0 < t' < T}}
\delta(t - t'),
\end{align}
where $\mathrm{sp}(i)$ is the (finite) set of spike times of neuron
$i$. We treat $s$ as the microstate of the system: the entire
spatiotemporal pattern of activity, not its instantaneous slice.

Given \(\mathcal G\), we associate with each microstate \(s\) an intrinsic activation field: the field that synaptic transmission triggered by \(s\) would generate in the absence of influences external to \(\mathcal G\). In the linear-response realization
this field can be specified as:
\begin{align}
V^{\mathrm{int}}_j(t) =
\left( \sum_{i \in \mathrm{pre}(j)} W_{ij}\, s_i(t - \Delta_{ij})
\right) * \eta(t),
\label{eq:Vint_general}
\end{align}
where \(\eta(t)\) is a causal response kernel.  In simple membrane-based realizations, such as the leaky
integrate-and-fire (IAF) realization introduced in
Sec.~\ref{sec:IAF_realization}, the endogenous field \(V^{\rm int}\)  can be identified with an
intrinsic voltage response obtained by filtering presynaptic spike
events. More generally, \(V^{\rm int}\) should be understood as an
effective activation field, and the linear convolution above may be
replaced by a more general response functional.

The central object of the framework is an \emph{energy functional}
$E(s)$ that
quantifies the consistency between a spike configuration and the
endogenous activation field generated by that same configuration:
\begin{align}
E(s) = -\sum_j \int dt\, s_j(t)\,
\mathcal{T}\!\left(V^{\mathrm{int}}_j(t)\right),
\label{eq:E_general}
\end{align}
where \(\mathcal T : \mathbb R \to \mathbb R_{\ge0}\) is a
possibly nonlinear thresholding function. It assigns a
non-negative energetic weight to the endogenous activation field: this
weight is large when the field favours a spike and small, or zero in the
hard-threshold case, when the field does not support spiking. With this
sign convention, a spike lowers the energy only to the extent that it
occurs at a time of endogenous support.

This energetic ordering should be understood relative to an admissible
class of spike trains. Without a constraint on the overall activity
level, adding further supported spikes can lower the energy. In the
canonical ensemble used below, however, the total spike count is fixed;
within such a fixed-activity model, lower-energy configurations are
those in which the prescribed spike events are better aligned with the
endogenous activation generated by the same configuration.\footnote{The construction is not specific to the IAF model. It only requires an
endogenous activation variable associated with each spike configuration
and a scalar support function \(\mathcal T\) assigning energetic weight
to that variable. In membrane-based realizations this variable may be a
subthreshold voltage followed by thresholding; in reduced realizations it
is more generally an effective activation field. The IAF case is treated
in Sec.~\ref{sec:IAF_realization}}.
 Note that, using the standard convolution convention
\[
    (f*\eta)(t)=\int d\tau\, f(\tau)\eta(t-\tau),
\]
Eq.~\eqref{eq:Vint_general} contains contributions of the form
\[
    (s_i(\cdot-\Delta_{ij})*\eta)(t)
    =
    \sum_{t_a\in{\rm sp}(i)}
    \eta(t-t_a-\Delta_{ij}) .
\]
Thus, for a causal response kernel, \(\eta(\tau)=0\) for
\(\tau<0\), and non-negative delays \(\Delta_{ij}\ge0\), the
intrinsic field of neuron \(j\) at time \(t\) receives contributions only
from presynaptic spikes whose delayed arrival times satisfy
\(t_a+\Delta_{ij}\le t\). The consistency measured by
Eq.~\eqref{eq:E_general} is therefore causal: a spike of neuron \(j\) is
energetically supported only by earlier, or in the zero-delay idealization
simultaneous, presynaptic events in the same spike-train configuration.

The remaining freedom in the network's spiking activity, that is, the
deviation of empirical spikes from the endogenously predicted ones, is
attributed to influences not represented explicitly by
\(\mathcal G\) or by the chosen response model, including external
synaptic input, channel noise, and unresolved intrinsic heterogeneity.  We collectively model this residual influence as a stochastic
heat bath. Applying Jaynes' maximum entropy principle \cite{Jaynes1957} with a
constraint on the mean energy \(\langle E\rangle\), we obtain the
least-biased distribution over admissible microstates in the Boltzmann
form
\begin{align}
P(s) &= Z^{-1} e^{-\beta E(s)}, \qquad s \in \mathcal{S},
\label{eq:Boltzmann} \\
Z &= \sum_{s \in \mathcal{S}} e^{-\beta E(s)},
\label{eq:Z}
\end{align}
where the sum runs over a prescribed set $\mathcal{S}$ of
\emph{admissible} microstates. The energy (\ref{eq:E_general}) is
bounded from below: each spike contributes at most a fixed negative amount
$-\mathcal{T}_{\max}$ to $E$, and the total spike count in a finite
window is itself bounded. In practice, boundedness is ensured by working at finite temporal
resolution, by imposing an absolute refractory period, i.e. a minimal
allowed interval between successive spikes of the same neuron, or by
fixing the total spike count. The ensemble (\ref{eq:Boltzmann})--(\ref{eq:Z}) is
therefore well-defined irrespective of $\mathcal{S}$. The role of
$\mathcal{S}$ is, instead, twofold: to encode hard biological
constraints (refractoriness, temporal resolution), and to disentangle
control parameters. In an unconstrained ensemble the mean spike count
$\langle n \rangle$ is itself a non-trivial function of $\beta$, so
that $\beta$ alone cannot serve as a clean measure of the external
drive at fixed firing rate. Restricting \(\mathcal S\) to configurations with a prescribed total
number of spikes,
\[
    n(s)=\sum_i\int_0^T dt\,s_i(t),
\]
separates the overall activity level from the parameter \(\beta\), which
controls the statistical weight of endogenous structure relative to
fluctuations. We adopt this canonical convention, with fixed \(n(s)\),
throughout the paper. A grand-canonical extension, in which a
chemical-potential-like Lagrange multiplier fixes \(\langle n\rangle\)
rather than \(n\) itself, is possible but is not used below.

The inverse temperature $\beta$ controls the relative weight of the
endogenous structure and the residual fluctuations: in the limit
$\beta \to \infty$ the ensemble concentrates on microstates maximally
consistent with the endogenous dynamics; in the limit $\beta \to 0$
all admissible microstates become equiprobable. We accordingly
identify $\beta^{-1}$ with the strength of the external drive --- the
``temperature'' of the neural heat bath.\footnote{The ensemble
(\ref{eq:Boltzmann}) is, by construction, a stationary distribution
over entire spike trains and not a temporal evolution rule. The time
variable $t$ within $s$ plays the role of an additional dimension,
with the caveat that interactions are causal:
$V^{\mathrm{int}}_i(t)$ depends only on spikes at times $t' < t$.}

\subsubsection*{Macrostates and free energy}

It is convenient to coarse-grain the ensemble by grouping microstates
that share a common energy. A \emph{macrostate} $m$ is an
equivalence class of admissible microstates with energy $E(m)$, and
its multiplicity is
\begin{align}
\Gamma(m) = \big|\{ s \in \mathcal{S} : E(s) = E(m) \}\big|.
\end{align}
The partition function (\ref{eq:Z}) decomposes as
\begin{align}
Z = \sum_m \Gamma(m)\, e^{-\beta E(m)} = \sum_m e^{-\beta F(m)},
\end{align}
with the macrostate free energy
\begin{align}
F(m) = E(m) - \frac{H(m)}{\beta}, \qquad
H(m) = \log \Gamma(m).
\label{eq:F}
\end{align}
The probability of macrostate $m$ is $P(m) = Z^{-1} e^{-\beta F(m)}$,
and the most probable macrostate is the global minimizer of $F$.
The free energy \(F\) in Eq.~\eqref{eq:F} is the central object of the
framework. In what follows, activity regimes are characterized by the
macrostates that minimize \(F\). Transitions between regimes are then
understood as qualitative changes, or finite-size crossovers, in the
properties of these minimizing macrostates.

\subsubsection*{Relation to existing statistical-mechanical formulations}

The present framework shares a maximum-entropy foundation with pairwise
models of spike data~\cite{Schneidman2006, Tkacik2013, Cocco2009, Roudi2009}: in both cases, Jaynes' principle leads to
a Boltzmann distribution over neural activity patterns. Three differences are worth emphasising. First, the
microstate. In those models the microstate is an instantaneous
binary vector $\sigma_t \in \{0,1\}^N$ over a single time bin, and
the ensemble lives on instantaneous configurations; here the
microstate is the full spatiotemporal spike train $s$ over $[0,T]$,
and the ensemble lives on trajectories. Second, the constraints.
Pairwise MaxEnt models constrain measured firing rates and pairwise
correlations, yielding an Ising-like Hamiltonian whose couplings are
\emph{inferred} from data; here the energy (\ref{eq:E_general}) is
\emph{constructed} from a known graph $\mathcal{G}$ and single-neuron
rule, while the inverse problem in Sec.~\ref{sec:conn}
treats connectivity parameters and a separate inference-precision
parameter as objects of estimation. Third, the role of time. The energy
(\ref{eq:E_general}) couples spikes across time through the causal
kernel $\eta$ and the delays $\Delta_{ij}$; the construction is
therefore closer in spirit to the generating-functional /
path-integral formulations of stochastic neural networks
\cite{MartinSiggiaRose1973, BuiceCowan2007, Crisanti1988}, with
$\beta$ playing a role analogous to a noise amplitude rather than a
true thermodynamic temperature.

\subsection{IAF realization}
\label{sec:IAF_realization}

To make the framework concrete, we now specialise the kernel $\eta$
and the threshold operator $\mathcal{T}$ to the leaky
integrate-and-fire (IAF) model, originally introduced by Lapicque
\cite{Lapicque1907, BrunelVanRossum2007} and extensively studied
since; for comprehensive modern treatments and reviews see
\cite{Tuckwell1988, Burkitt2006a, Burkitt2006b, Gerstner2014}. A single neuron in this model
evolves according to
\begin{align}
\frac{dV}{dt} = -\frac{V}{\tau} +
\kappa\!\left(I^{\mathrm{int}} + I^{\mathrm{ext}}\right),
\label{eq:LIF}
\end{align}
where $V$ is the membrane voltage, $\tau$ the membrane time constant,
and $\kappa$ a constant of dimension $\textrm{Ohm}/\mathrm{s}$ that
converts the synaptic current into a voltage rate (e.g.,
$\kappa = R_m / \tau$ for a passive membrane of resistance $R_m$).
The intrinsic current $I^{\mathrm{int}}$ arises from synaptic
transmission within the population; $I^{\mathrm{ext}}$ collectively
represents synaptic input from outside the population and any other
influences not encoded in $\mathcal{G}$. We assume that the
statistical properties of $I^{\mathrm{ext}}$ are stationary in time.

Intrinsic synaptic interactions are taken to be impulsive: when
neuron $i$ fires at time $t_i^{\mathrm{sp}}$, the intrinsic current
in neuron $j$ acquires an instantaneous contribution
\begin{align}
I^{\mathrm{int}}_j \to I^{\mathrm{int}}_j +
W_{ij}\, \delta\!\left(t - t_i^{\mathrm{sp}} - \Delta_{ij}\right),
\end{align}
so that all postsynaptic charge is delivered in a single
pulse.\footnote{Replacing $\delta$ by an arbitrary postsynaptic
current kernel recovers the Spike Response Model \cite{Gerstner95}.} In the
deterministic version of the IAF model, spiking is governed by the
threshold rule
\begin{align}
\text{if } V_i(t) > \theta:\quad
\mathrm{sp}(i) \to \mathrm{sp}(i) \cup \{t\},\quad
V_i(t) \to V_{\mathrm{reset}}.
\end{align}
Integrating (\ref{eq:LIF}) with the impulsive synaptic input above
yields the unit-pulse response kernel
\begin{align}
\eta(t) = \kappa\, \Theta(t)\, \exp(-t/\tau),
\label{eq:eta_IAF}
\end{align}
which together with (\ref{eq:Vint_general}) defines the intrinsic
voltage response $V^{\mathrm{int}}_j(t)$ of the IAF realization.

For the threshold operator we take the Heaviside form
\begin{align}
\mathcal{T}(V) = \Theta(V - \theta),
\label{eq:T_IAF}
\end{align}
so that the energy functional (\ref{eq:E_general}) reduces to
\begin{align}
E(s) = -\sum_i \int dt\, s_i(t)\,
\Theta\!\left(V^{\mathrm{int}}_i(t) - \theta\right),
\label{eq:E_IAF}
\end{align}
i.e., minus the number of spikes whose timing is consistent with a
suprathreshold intrinsic potential. 
The threshold \(\theta\) acts as a structural control parameter of the
energy functional. In the binary-weight setting it fixes the effective
number of approximately coincident presynaptic events required for a
postsynaptic spike to be counted as endogenously supported. Its numerical
value is therefore scale-dependent: in biologically sized networks one
would expect a substantially larger effective threshold, whereas in the
small networks used in our Monte Carlo simulations we use lower values,
typically \(\theta\sim 3\), to obtain comparable levels of endogenous
support at finite connectivity.
With (\ref{eq:eta_IAF}) and
(\ref{eq:T_IAF}) inserted into (\ref{eq:Boltzmann})--(\ref{eq:Z}),
the abstract ensemble of Sec.~\ref{sec:abstract}  becomes an IAF-based spike-train ensemble. This
realization provides the biological reference point for the general
construction, while the analytical and numerical treatments below also
use reduced response operators when this is sufficient to isolate the
statistical-mechanical structure of the model. The
admissible set $\mathcal{S}$ will, throughout, enforce a fixed mean
firing rate and an absolute refractory period; further constraints
(e.g., minimal temporal resolution) are introduced as needed. A
smoothed sigmoidal version of $\mathcal{T}$, used in Sec.~\ref{sec:meanfield} to obtain
a tractable mean-field limit, is biologically natural insofar as it
accounts for the modulating (rather than gating) effect of
subthreshold voltage fluctuations on spike probability, and brings
the construction into formal closeness with stochastic IAF models in
the soft-threshold  \cite{SacerdoteRev11}.

\subsection{Discrete transfer-operator realization used in Monte Carlo}
\label{sec:discrete_operator_realization}

The IAF construction of Sec.~\ref{sec:IAF_realization} provides a
biologically motivated realization of the abstract ensemble introduced in
Sec.~\ref{sec:abstract}. The Monte Carlo simulations used below employ a
further reduced, discrete realization of the same ensemble. This
realization is not intended to represent an explicit time integration of
the IAF membrane equation. Instead, it implements the endogenous network
response by means of a linear transfer operator acting directly on binned
spike trains.

For the Monte Carlo realization, we use a finite-bin version of the
event-based spike train introduced in Sec.~\ref{sec:abstract}. The
observation window is represented by \(K\) time bins, and the spike train
\(s_i(t)\) is replaced by a binary raster
\[
    n_{i\ell}\in\{0,1\},
    \qquad
    i=1,\ldots,N,
    \qquad
    \ell=0,\ldots,K-1,
\]
where \(n_{i\ell}=1\) denotes a spike of neuron \(i\) in time bin \(\ell\).
The Monte Carlo microstate is therefore
\[
    n \in \{0,1\}^{N\times K}.
\]
The admissible set \(\mathcal S_K\) is restricted by two hard constraints:
a fixed total number of spikes,
\[
    \sum_{i=1}^{N}\sum_{\ell=0}^{K-1} n_{i\ell}=M,
\]
and an absolute refractory constraint excluding spike pairs of the same
neuron separated by fewer than \(r_{\rm ref}\) time bins. This is the
discrete canonical counterpart of the fixed-activity ensemble discussed
in Sec.~\ref{sec:abstract}.

For a fixed graph \(G=(V,E,W,\Delta)\), the intrinsic response is
represented by a discrete transfer operator \(\mathcal B_G\) acting on
the binary raster,
\[
    u=\mathcal B_G n .
\]
The field \(u\) is interpreted as the effective endogenous activation
generated by the spike train \(n\) itself. It plays the role of the
intrinsic response field \(V^{\rm int}\) in the abstract formulation, but
it is not obtained by solving a membrane equation. Rather, it is a
reduced activation field encoding delayed propagation through the network
and temporal filtering by an effective response kernel.

In the Monte Carlo simulations reported here we use homogeneous
excitatory weights, \(W_{ij}=1\) for \((i,j)\in E\), and \(W_{ij}=0\)
otherwise. Writing \(A_{ij}\in\{0,1\}\) for the adjacency matrix, the
operator \(\mathcal B_G\) is implemented in the temporal Fourier
representation. The explicit construction, including the causal delay
phase, the response kernel, and the inverse transform back to the time
domain, is given in Appendix~\ref{app:discrete_transfer_operator}.

The discrete energy functional is then defined in the time domain as
\begin{equation}
    E_K(n)
    =
    -
    \sum_{i=1}^{N}
    \sum_{\ell=0}^{K-1}
    n_{i\ell}\,
    \phi(u_{i\ell})
    =
    -
    \sum_{i=1}^{N}
    \sum_{\ell=0}^{K-1}
    n_{i\ell}\,
    \phi\!\left[
        (\mathcal B_G n)_{i\ell}
    \right],
\label{eq:discrete_mc_energy}
\end{equation}
where \(\phi:\mathbb R\rightarrow\mathbb R\) is a soft or hard
thresholding nonlinearity. The simulations reported in Fig.~\ref{fig:mc_mean_field}
use the hard-threshold choice
\[
    \phi(x)=\Theta(x-\theta),
    \qquad
    \theta=3.
\]
Other nonlinearities, such as smooth sigmoidal or rectified responses,
can be implemented within the same operator-based construction, but are
not analysed in the numerical phase map presented below. Equation
\eqref{eq:discrete_mc_energy} is the discrete counterpart of
Eq.~\eqref{eq:E_general}: a spike lowers the energy when it occurs at a
site where the endogenously generated activation field is large.

This reduced operator-based realization belongs to the same general class
of trajectory ensembles as the IAF realization, but replaces the
continuous membrane-voltage dynamics by an effective activation field
computed directly from the spike train. Its purpose is to isolate the
statistical-mechanical consequences of delayed network transfer,
refractory admissibility constraints, and the competition between
energetic self-consistency and configurational entropy.

\section{Mean-field reduction of the trajectory ensemble}
\label{sec:meanfield}

\subsection{Linear-response mean-field approximation and population-rate macrostates}

In order to obtain an analytically tractable description of the
trajectory ensemble, we now introduce a mean-field reduction based on a
linear-response approximation. The purpose of this reduction is not to
provide a quantitatively exact theory over the entire phase diagram, but
to construct an analytically transparent reference theory for the
macroscopic organization of spike histories. 
In particular, the
linearized theory derived below should be interpreted as an envelope for
the microscopic Monte Carlo ensemble once finite-size effects, quenched
connectivity disorder, refractory constraints, and nonlinear thresholding
are restored.

We first replace the hard threshold by a smooth sigmoidal response,
which preserves the qualitative role of thresholding while allowing a
controlled small-field expansion. Thus, instead of the Heaviside response
in Eq.~\eqref{eq:E_IAF}, we consider
\begin{align}
\label{eq:e_tanh}
E(s)
=
-\sum_i\int dt\;s_i(t)
\tanh\!\left(\frac{V^{\rm int}_i(t)}{\gamma}\right),
\end{align}
where \(\gamma\) is a sensitivity scale setting the voltage range over
which the response crosses over from subthreshold to saturated. We keep
\(\gamma\) explicit in order to track how the microscopic inverse
temperature is rescaled in the linearized mean-field theory.

The choice of a sigmoidal threshold is biologically motivated by the
fact that subthreshold voltage fluctuations around resting potential
(which in our case equals \(0\)) evoked by intrinsic transmission can
influence the likelihood of spiking.\footnote{This soft-threshold
formulation is closely related to stochastic IAF models, where
subthreshold fluctuations modulate the probability of spike generation.}

We then restrict the analytic mean-field calculation to the small-field
regime
\[
    |V^{\rm int}_i(t)|/\gamma \ll 1,
\]
in which
\begin{align}
\label{eq:VtanhV}
\tanh\!\left(\frac{V^{\rm int}_i(t)}{\gamma}\right)
\approx
\frac{V^{\rm int}_i(t)}{\gamma}.
\end{align}
This is the linear-response approximation used throughout
Sec.~\ref{sec:phases}. It is most accurate when the endogenous voltage
generated by the recurrent network remains small compared with the
sensitivity scale \(\gamma\). Outside this regime the linear theory overestimates the energetic gain
associated with large endogenous fields. The resulting phase boundaries
should therefore be interpreted cautiously.  A precise outer-estimate
statement is established below only for the point-collapse boundary
relative to the uniform macrostate.

We now assume a large homogeneous excitatory population with sparse
random connectivity, \(p_{\rm conn}\sim N^{-1}\), so that the expected
number of presynaptic inputs per neuron remains finite as \(N\) grows.
The corresponding finite factor \(p_{\rm conn}N\) is absorbed into the
effective mean synaptic weight \(\langle W\rangle\).
We denote by
\[
    \alpha
    =
    \frac{1}{N}
    \sum_{i=1}^{N}|{\rm sp}(i)|
\]
the average number of spikes per neuron in the observation window. Since in the
distributional notation we have \(s_i(t)=\sum_{t_a\in{\rm sp}(i)}
\delta(t-t_a)\), $\alpha$ can be equivalently rephrased as:
\[
    \alpha
    =
    \frac{1}{N}
    \sum_{i=1}^{N}
    \int_0^T dt\,s_i(t),
\]
where the integral denotes the total mass of the spike-train measure. In
the canonical ensemble used throughout the paper, \(\alpha\) is fixed.

We introduce coarse grained population rate macrostates by smoothing the
population spike train over a mesoscopic temporal scale \(\tau\):
\begin{align}
\label{eq:rho_mf}
\rho_\tau(t)
=
\frac{1}{\alpha N}
\int dt'\;
K_\tau(t-t')
\sum_i s_i(t'),
\end{align}
where \(K_\tau\ge0\) is a normalized smoothing kernel,
\(\int dt\,K_\tau(t)=1\). The scale \(\tau\) is taken to be larger than
the microscopic temporal resolution but smaller than the macroscopic
scale over which the population rate varies. In the following we write
\(\rho(t)\) for this smooth mean-field density.  Assuming periodic boundary conditions on
\([0,T]\), or equivalently a smoothing kernel normalized on the
observation window, Eq.~\eqref{eq:rho_mf} gives
\[
    \int_0^T dt\,\rho_\tau(t)=1.
\]
Thus \(\rho(t)\) is the temporal probability density of a spike drawn
from the population spike history. Although \(\rho(t)\) is a function of time, it should not be interpreted
as an instantaneous dynamical state of the population. It is a
trajectory-level macrostate: a coarse-grained temporal profile extracted
from the complete spike history \(s\) over the observation window
\([0,T]\). Thus the mean-field reduction replaces the full
spatiotemporal spike-train microstate by a lower-dimensional collective
description of how the spikes of the entire trajectory are distributed
in time.

\subsection{Effective delay kernel and linearized mean-field free energy}

Next, we introduce the delayed average voltage response kernel
\begin{align}
\Omega(t)
=
\left[
\left\langle W_{ij}\mid \Delta_{ij}=t\right\rangle
P_\Delta(t)
\right]*\eta(t),
\end{align}
where \(P_\Delta(t)\) is the normalized distribution of synaptic delays,
\(\eta\) is the voltage response kernel to a unit current pulse, and
\(*\) denotes convolution. Assuming independence of \(W_{ij}\) and
\(\Delta_{ij}\), we write
\begin{align}
\Omega(t)
=
\langle W\rangle
\left(P_\Delta*\eta\right)(t).
\end{align}
It is useful to separate the overall coupling scale from the shape of the
delay-response kernel. Absorbing the normalization of
\(P_\Delta*\eta\) into the effective mean synaptic weight, we define a
normalized kernel
\[
    \widehat\Omega(t)
    =
    \left(P_\Delta*\eta\right)(t),
    \qquad
    \int dt\,\widehat\Omega(t)=1,
\]
so that
\[
    \Omega(t)=\langle W\rangle\widehat\Omega(t).
\]
Here the normalization of \(P_\Delta*\eta\), if different from unity, is
absorbed into the effective mean synaptic weight \(\langle W\rangle\).

The average intrinsic voltage response in a single neuron at time \(t\)
is then
\begin{align}
\label{eq:av_voltage}
\left\langle V^{\rm int}(t)\right\rangle
=
\alpha\langle W\rangle
\int dt'\;
\widehat\Omega(t-t')\rho(t').
\end{align}
This step is an annealed mean-field replacement of the quenched adjacency
and delay realization by its ensemble-averaged delay-response kernel. The
microscopic Monte Carlo simulations in Sec.~\ref{sec:monte_carlo}
restore fixed graph and delay realizations and therefore include
realization-specific effects absent from the annealed mean-field
description.
Using the linear response in Eq.~\eqref{eq:VtanhV}, the total
mean-field energy of the macrostate becomes
\begin{align}
\label{eq:mf_energy}
E_{\rm lin}^{\rm tot}[\rho]
=
-\frac{\alpha^2 N\langle W\rangle}{\gamma}
\iint dt\,dt'\;
\rho(t)\widehat\Omega(t-t')\rho(t').
\end{align}
Equations~\eqref{eq:av_voltage} and~\eqref{eq:mf_energy} can also be
expressed in terms of the autocorrelation of \(\rho\), a form used later
in Sec.~\ref{sec:conn}.

The entropy of the normalized spike density \(\rho(t)\) is
\begin{align}
H[\rho]
=
-\int dt\;\rho(t)\log\rho(t).
\end{align}
The number of microscopic spike histories realizing a smooth density
\(\rho\) scales as
\[
    \Gamma[\rho]
    \sim
    \exp\!\left(\alpha N H[\rho]\right).
\]
Thus the linearized free-energy density per spike, written in terms of
the microscopic inverse-temperature parameter \(\beta\) of the canonical
trajectory ensemble, is
\begin{align}
\label{eq:f_mf_beta}
f_{\rm lin}[\rho]
=
\frac{E_{\rm lin}^{\rm tot}[\rho]}{\alpha N}
-
\frac{H[\rho]}{\beta}.
\end{align}
Introducing
\[
    Q[\rho]
    =
    \iint dt\,dt'\;
    \rho(t)\widehat\Omega(t-t')\rho(t'),
\]
we can write
\[
    f_{\rm lin}[\rho]
    =
    -C Q[\rho]
    -
    \frac{H[\rho]}{\beta},
    \qquad
    C
    =
    \frac{\alpha\langle W\rangle}{\gamma}.
\]
Since multiplication of the free energy by a positive constant does not
change its minimizers, we divide by \(C\) and obtain the equivalent
dimensionless mean-field functional
\begin{align}
\label{eq:F_mf}
\mathcal F_{\rm lin}[\rho]
=
E_{\rm MF}[\rho]
-
\frac{H[\rho]}{\beta_{\rm MF}},
\end{align}
where
\begin{align}
\label{eq:E_mf_rescaled}
E_{\rm MF}[\rho]
&=
-Q[\rho]
=
-\iint dt\,dt'\;
\rho(t)\widehat\Omega(t-t')\rho(t'),
\\
\label{eq:beta_mf}
\beta_{\rm MF}
&=
\beta C
=
\frac{\beta\alpha\langle W\rangle}{\gamma}.
\end{align}
Here \(\beta\) remains the inverse-temperature parameter of the original
canonical trajectory ensemble, whereas \(\beta_{\rm MF}\) is the
effective inverse temperature appearing in the rescaled linearized
mean-field functional. Thus, in the linearized mean-field theory,
\(\beta\), \(\alpha\), \(\langle W\rangle\), and \(\gamma\) enter the
phase-boundary calculation only through the combination
\(\beta_{\rm MF}\). This reflects a parameter degeneracy of the
linear-response approximation and does not imply that the microscopic
parameters are physically indistinguishable in the full nonlinear
ensemble.

This formulation also clarifies the status of the linearized phase
boundaries discussed below. Using Eq.~\eqref{eq:av_voltage}, the
dimensionless mean endogenous field entering the sigmoid is
\begin{eqnarray}
\label{eq:u_rho}
u_\rho(t)
=
\frac{\left\langle V^{\rm int}(t)\right\rangle}{\gamma}
&=&
\frac{\alpha\langle W\rangle}{\gamma}
\int dt'\;\widehat\Omega(t-t')\rho(t')\nonumber\\
&=&
\frac{\alpha\langle W\rangle}{\gamma}
(\widehat\Omega*\rho)(t).
\end{eqnarray}
For the same macrostate \(\rho\), the unrescaled linearized and
saturating mean-field energies are therefore
\begin{align}
E_{\rm lin}^{\rm tot}[\rho]
&=
-\alpha N
\int dt\;\rho(t)u_\rho(t),
\\
E_{\rm sat}^{\rm tot}[\rho]
&=
-\alpha N
\int dt\;\rho(t)\tanh u_\rho(t).
\end{align}
For purely excitatory kernels, \(\widehat\Omega\ge0\), and non-negative
densities, \(u_\rho(t)\ge0\). Since \(u-\tanh u\ge0\) for \(u\ge0\), one
has
\[
    E_{\rm sat}^{\rm tot}[\rho]
    -
    E_{\rm lin}^{\rm tot}[\rho]
    =
    \alpha N
    \int dt\;\rho(t)
    \bigl[u_\rho(t)-\tanh u_\rho(t)\bigr]
    \ge0 .
\]
Thus the linearized theory assigns an energy that is at least as
favourable as the saturating theory to any fixed macrostate. This pointwise
comparison does not by itself determine the nesting of phase boundaries,
which depends on energy gaps between competing macrostates. In Sec.~\ref{sec:phases}, this comparison is made explicitly for the
idealized point-collapsed density \(\rho_\delta\) and the uniform density
\(\rho_0\). For periodic-collapsed macrostates, the same saturation mechanism is
expected to act qualitatively, although the precise comparison depends
on the window weights and on the shape of the delay kernel.

In the analytic calculations below we focus on the simplest kernel shape
by taking
\begin{align}
\label{eq:trivkernel}
\eta(t)=\delta(t),
\end{align}
so that \(\widehat\Omega(t)=P_\Delta(t)\). The overall multiplicative strength of the recurrent interaction is then
contained in
\(\beta_{\rm MF}=\beta\alpha\langle W\rangle/\gamma\), while
\(\widehat\Omega(t)=P_\Delta(t)\) describes only the shape of the delay
kernel. This choice isolates the role of
the delay distribution itself. It also permits instantaneous interactions
when the support of \(P_\Delta\) reaches zero; this idealization will be
discussed below.

In the following sections we evaluate \(\mathcal F_{\rm lin}\) for a
range of idealized densities \(\rho\) and explore their dependence on the
effective mean-field inverse temperature \(\beta_{\rm MF}\) and on the
shape of the delay distribution \(P_\Delta(t)\). We consider delay
distributions on \(\mathbb R^+\) characterized by their mean
\(\mu_\Delta\) and standard deviation \(\sigma_\Delta\). Two useful
reference cases are a narrow-delay Gaussian approximation,
\begin{align}
\label{eq:Gaussian}
P_\Delta(t)
&\simeq
\mathcal N(\mu_\Delta,\sigma_\Delta),
\qquad
\sigma_\Delta\ll\mu_\Delta,
\end{align}
and the exponential distribution,
\begin{align}
\label{eq:exponential_dist}
P_\Delta(t)
&=
\mu_\Delta^{-1}e^{-t/\mu_\Delta},
\qquad
\sigma_\Delta=\mu_\Delta.
\end{align}
Here Eq.~\eqref{eq:Gaussian} denotes a Gaussian approximation to almost
fixed delays, while Eq.~\eqref{eq:exponential_dist} is the
maximum-entropy distribution on \(\mathbb R^+\) with prescribed mean.

\section{Mean-field phases of excitatory delayed networks}
\label{sec:phases}

Throughout this section we analyze the continuous-time mean-field
functional derived in Sec.~\ref{sec:meanfield}. Thus \(t\in[0,T]\) is a
continuous time variable and \(\rho(t)\) is a normalized population-rate
density, in contrast to the discrete spike trains used in the
microscopic ensemble and in the Monte Carlo simulations. The microscopic
temporal binning and refractory constraints re-enter only in
Sec.~\ref{sec:monte_carlo}. For notational economy, we write
\(\Omega\equiv\widehat\Omega\) for the normalized delay-response kernel
introduced in Sec.~\ref{sec:meanfield}. All energies below are the
rescaled dimensionless mean-field energies \(E_{\rm MF}\) entering
\[
    \mathcal F_{\rm lin}[\rho]
    =
    E_{\rm MF}[\rho]
    -
    \frac{H[\rho]}{\beta_{\rm MF}},
\]
where \(\beta_{\rm MF}\) is the effective mean-field inverse temperature
defined in Eq.~\eqref{eq:beta_mf}. To lighten notation, we write \(E(\rho)\) instead of \(E_{\rm MF}[\rho]\)
in this section; it should not be confused with the microscopic energy
\(E(s)\) of the original spike-train ensemble.

\subsection{Uniform state}

Let us consider normalized population spiking-rate densities over a
finite time interval \([0,T]\) with periodic boundary conditions
(somewhat unrealistic but acceptable when \(T\gg\mu_{\Delta}\)). A
uniform density \(\rho_0=1/T\) maximizes entropy and will always be the
preferred macrostate of an uncoupled system. Its dimensionless
mean-field energy and entropy are
\begin{align}
E(\rho_0) &= - \frac{1}{T},\\
H(\rho_0) &= \log(T),
\end{align}
and hence
\[
    \mathcal F_{\rm lin}[\rho_0]
    =
    -\frac{1}{T}
    -
    \frac{\log T}{\beta_{\rm MF}}.
\]
In the presence of coupling, however, it has to compete with other
macrostates.

\subsection{Semi-periodic states}

Let us now consider a family of semi-periodic  densities
\(\rho_\Delta\) of the form:
\begin{equation}
\label{eq:periodic_dens1}
    \rho_\Delta(t,\phi)
    =
\Sha_\Delta(t,\phi,\lambda^0,\ldots,\lambda^K)*B(t),
\end{equation}
with
\begin{equation}
\label{eq:periodic_dens2}\Sha_\Delta(t,\phi,\lambda^0,\ldots,\lambda^K)
    =
\sum_{k=0}^{K}\lambda^k\delta(t-k\Delta+\phi),
    \,\,
    \sum_{k=0}^{K}\lambda^k=1 .
\end{equation}
We work on the periodic observation window and take
\(T=(K+1)\Delta\), or equivalently neglect the boundary remainder in the
regime \(T\gg\Delta\). The kernel \(B(t)\) is a fixed normalized
activity-window shape. Since an overall shift of \(B\) can be absorbed
into the phase \(\phi\), we may take \(B\) to be supported on an interval
\([0,b]\) with \(b<\Delta\). The phase \(\phi\) labels equivalent
translations of the same macrostate and will be dropped below. The
superscript in \(\lambda^k\) is an index labeling the weight of the
\(k\)-th activity window, not exponentiation.

The ansatz represents macrostates in which population activity is
organized into repeated activity windows separated by a characteristic
period \(\Delta\). To calculate the energy of these macrostates, define:
\begin{align}
\label{eq:omega_k}
\omega^k(\Delta)
=
\int dt\;
B(t-k\Delta)
\left[B(t)*\Omega(t)\right].
\end{align}
The \(k\)-th term of this sequence represents the interaction between
cycles that are \(k\) periods apart in the oscillation
(\(\omega^0\) represents self-interaction). Since we assumed periodic
boundary conditions, we can express the energy as
\begin{align}
\label{eq:energy_lambdas}
E(\rho_{\Delta})
=
-\sum_{k=0}^{K}\lambda^k
\left(
\sum_{k'=k}^{\infty}
\lambda^{\operatorname{mod}(k',K+1)}
\omega^{(k'-k)}(\Delta)
\right),
\end{align}
which accounts for all possible cycle-cycle interactions.\footnote{Note
that the energy \eqref{eq:energy_lambdas} is invariant under shifting
the index \(k\rightarrow k+k'\), which, as these equivalent-energy
macrostates are separated by energy barriers, leads to spontaneous
symmetry breaking. The same applies to the phase \(\phi\) in
Eqs.~\eqref{eq:periodic_dens1}--\eqref{eq:periodic_dens2}.}

\begin{figure}
    \centering
    \includegraphics[width=\linewidth]{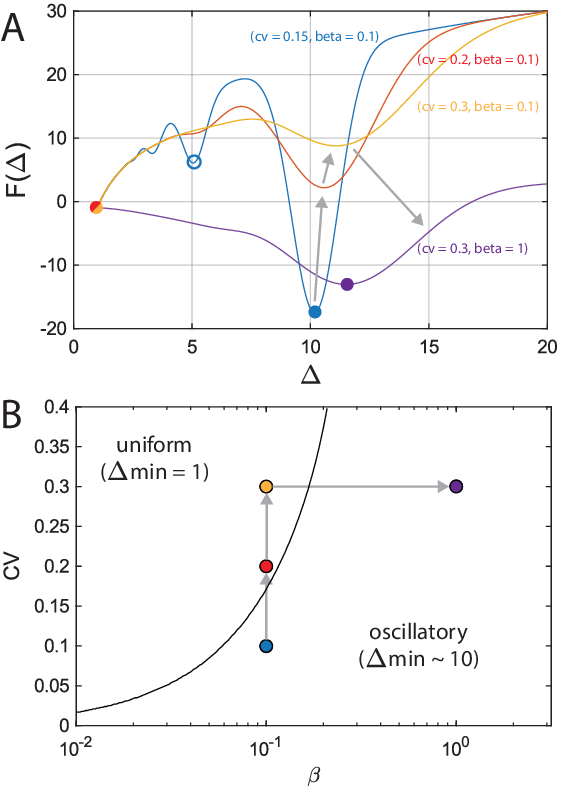}
    \caption{Representative examples of the free energy functional
    \(\mathcal F_{\rm lin}\) under the assumption \(B(t)=\delta(t)\).
    (A) With decreasing delay distribution coefficient of variability
    (CV), the free energy global minimum (purple and blue filled circles)
    becomes progressively deeper. Additional metastable minima also
    become more pronounced (open circle). A discontinuous transition
    occurs when the position of the global minimum crosses over to \(0\)
    with decreasing effective mean-field inverse temperature
    \(\beta_{\rm MF}\). (B) Parameter space positions corresponding to
    functions in (A). The line represents the phase transition boundary. Panel shows only the transition within the equal-weight
semi-periodic ansatz discussed in Sec.~\ref{sec:phases}\,B. It should not be read as
the full three-regime mean-field phase diagram of Fig.~\ref{fig:mc_mean_field}A,
which additionally includes the collapsed states analyzed in
Sec.~\ref{sec:phases}\,C.}
    \label{fig:figMC}
\end{figure}

Let us now focus on a sub-class of periodic densities where
\(\lambda^0=\lambda^1=\dots=\lambda^K=(K+1)^{-1}\sim\Delta/T\). These
densities are generally not energy-minimizing, as discussed in the next
section, but may serve as a good approximation close to a
uniform/periodic phase transition, since they maximize entropy in their
class:
\begin{align}
H(\rho_\Delta)
=
\log(K+1)+H(B)
\nonumber\\
\sim
\log(T)-\log(\Delta)+H(B),
\end{align}
where \(H(B)\) is the entropy of the kernel \(B\) and
\(H(B)\leq\log(\Delta)\). Note that \(\rho_{\Delta}\) behaves like a
discrete distribution. This leads to some problems, including violation
of the second Khinchin axiom by dense Dirac combs with very high entropy,
which can be addressed by bounding \(\Delta\) from below,
\(\Delta\in[\Delta^{\min},T]\). The lower limit \(\Delta^{\min}\) may
represent an absolute refractory period or, more generally, a minimal
resolvable temporal scale. Let us assume \(\Delta^{\min}=1\), so that
\(\lim_{\Delta\rightarrow\Delta^{\min}}H(\rho_\Delta)=H(\rho_0)\). With
all \(\lambda^k\) equal, the energy is now, up to a factor, just the sum
of Eq.~\eqref{eq:omega_k}:
\begin{align}
E(\rho_{\Delta})
=
-
\frac{\Delta}{T}
\sum_{k=0}^{\infty}
\omega^k(\Delta).
\end{align}
We can proceed by assuming \(B(t)=\delta(t)\). The sequence elements then
reduce to
\begin{align}
\omega^k(\Delta)
&=
\Omega(k\Delta),
\\[10pt]
\label{eq:energy_Omega}
E(\rho_{\Delta})
&=
-
\frac{\Delta}{T}
\sum_{k=0}^{\infty}
\Omega(k\Delta).
\end{align}
The energy \eqref{eq:energy_Omega} samples the delay kernel \(\Omega\)
at integer multiples of the candidate period \(\Delta\). It therefore
favours periods for which these sampled values overlap strongly with the
maxima of \(\Omega\) (Fig.~\ref{fig:figMC}A). Provided that
\(\Omega(t)\) has a single maximum at \(t^*\sim\mu_{\Delta}\), the energy
contains a series of progressively deeper local minima at integer
divisions \(\Delta=t^*/n\), \(n\in\mathbb N^+\), corresponding to
oscillation harmonics, and a global minimum located at \(t^*\), after
which it decays sharply to zero. In the limit \(\Delta\rightarrow0\), the
energy converges, up to a factor, to the integral of \(\Omega\). If the
only densities allowed in our ensemble were uniform and periodic, within
the sub-class discussed above, then the minimum-free-energy principle
would predict oscillatory behavior when
\begin{align}
\min_{\Delta}\mathcal F_{\rm lin}(\rho_\Delta)
<
\mathcal F_{\rm lin}(\rho_0),
\end{align}
which corresponds to
\begin{align}
\label{eq:periodic_transition_condition}
\frac{\Delta}{T}
\sum_{k=0}^{\infty}\omega^k(\Delta)
>
\frac{1}{T}
+
\frac{\log(\Delta)-H(B)}{\beta_{\rm MF}}.
\end{align}
We see that with decreasing \(\beta_{\rm MF}\), i.e. with weaker
effective selection by the linearized energy relative to entropy,
progressively more compact delay distributions are required to support a
delay-induced oscillation. Broader kernels \(B(t)\) may partially
compensate for this effect. In the limit \(\beta_{\rm MF}\to0\), the
uniform density dominates regardless of the shape of the delay
distribution. See Fig.~\ref{fig:figMC}A--B for an illustration of the
parameter dependence of \(\mathcal F_{\rm lin}\) and its minimum.

\subsection{Collapsed discharge-like states}

The quadratic form of the linearized mean-field energy favors sharply
peaked densities, reminiscent of neuronal population discharges
\cite{Menendez11, Huberfeld11, Buzsaki15}, because concentrated profiles
generate larger endogenous fields and are therefore over-rewarded by the
linear response. As discussed in Sec.~\ref{sec:meanfield},   this motivates
using the linearized collapse boundaries as outer estimates. The precise
outer-estimate statement is established below only for the
point-collapsed state relative to the uniform state; periodic-collapsed
states are treated as a qualitative extension of the same mechanism.

\subsubsection{Periodic collapse}

Let us start with periodic densities where, in this more general case,
we need to explicitly consider the weights \(\lambda^k\) of individual
activity windows in Eq.~\eqref{eq:energy_lambdas}. We will still assume
\(B(t)=\delta(t)\), which becomes a better approximation with increasing
\(\beta_{\rm MF}\), as the energy term increasingly favors temporally
concentrated densities at the cost of entropy. The entropy is then
\begin{align}
H(\rho_\Delta)
=
-\sum_{k=0}^{K}\lambda^{k}\log(\lambda^k).
\end{align}
As before, the energy typically exhibits a series of minima corresponding
to subsequent harmonics, independently of the exact choice of
\(\lambda^k\). In the case of unimodal \(\Omega\), however, the deepest
energy minimum is always associated with the fundamental period
\(\Delta=t^*\). Hence, if we assume that self-interaction is negligible
(\(\omega^0=0\)) and \(\Omega\) decays sufficiently fast, we can consider
only nearest-neighbor interactions and omit all terms except
\(\omega^1\). Under these assumptions, the energy
\eqref{eq:energy_lambdas} factorizes:
\begin{align}
\label{eq:energy_lambdas_factored}
E(\rho_{\Delta=t^*})
=
-\Omega(t^*)
\left[
\sum_{k=1}^{K}\lambda^{k-1}\lambda^k
\right].
\end{align}
To minimize energy we need to maximize the sum in square brackets under
the normalization constraint \(\sum_{k=0}^K\lambda^{k}=1\). Writing $a=\sum_{k\ \mathrm{even}}\lambda^{k}$ and $b=\sum_{k\ \mathrm{odd}}\lambda^{k}$,
so that $a+b=\sum_{k=0}^{K}\lambda^{k}=1$, we note that every term of the sum
$\sum_{k=1}^{K}\lambda^{k-1}\lambda^{k}$ pairs an even-indexed window with an
odd-indexed one. The sum is therefore bounded by the full bipartite product,
\begin{equation}
\sum_{k=1}^{K}\lambda^{k-1}\lambda^{k}\;\le\;ab\;\le\;\frac{(a+b)^{2}}{4}\;=\;\frac{1}{4},
\end{equation}
independently of the number of windows $K$, where the second inequality is the
arithmetic--geometric mean inequality. Equality requires $a=b=1/2$ together with
the vanishing of all non-adjacent even--odd products; this is realized by the
family:
\begin{equation}
    \label{eq:lam_family}
    \lambda^{0}=x,\ \lambda^{1}=1/2,\
\lambda^{2}=1/2-x,
\end{equation}
 with $x\in[0,1/2]$ and all other $\lambda^{k} = 0$ for $k\geq 3$. The maximizer is thus degenerate, and the
energy-minimizing density concentrates on two or three consecutive activity
windows. The
term in brackets in Eq.~\eqref{eq:energy_lambdas_factored} is then
exactly \(1/4\). The energy-minimizing density is therefore expected to
collapse to only two or three cycles of an oscillatory neuronal discharge-like state at
frequency \(1/t^*\).

In the case of broader \(\Omega\), we can no longer ignore the
self-interaction term \(\omega^0\). If we continue to neglect all terms
beyond \(\omega^1\), the energy is now
\begin{align}
E(\rho_{\Delta})
=
-\omega^0
\left[
\sum_{k=0}^{K}(\lambda^k)^2
\right]
-
\omega^1
\left[
\sum_{k=1}^{K}\lambda^{k-1}\lambda^k
\right].
\end{align}
For \(\lambda\) belonging to the family \eqref{eq:lam_family}, the first
term in brackets is no more than \(0.5\), while it is maximal when
\(\lambda^0=1\) and all other \(\lambda^k=0\) for \(k\geq1\). We
therefore expect oscillatory collapse to be energy-minimizing only as
long as
\[
    \omega^1>2\omega^0.
\]
For such periodic-collapsed states, saturation acts through the same
pointwise deficit \(u-\tanh u\) discussed in Sec.~\ref{sec:meanfield},
but the ordering of the corresponding integrated deficits relative to
the uniform state depends on the detailed window weights and kernel
shape. We therefore use the linearized boundary as a qualitative outer
estimate rather than as a sharp saturating-theory prediction.
\subsubsection{Point collapse}

Let us now consider the exponential delay distribution
\eqref{eq:exponential_dist}, which represents the broad,
maximum-entropy case on \(\mathbb R^+\) at fixed mean. It can be shown
that now \(\omega^0>\omega^1\) for all choices of \(B\) and \(\Delta\).
Indeed, the rescaled mean-field energy is
\begin{align}
E(\rho_{\delta})
=
-\Omega(0)
=
-\frac{1}{\mu_\Delta},
\end{align}
minimized trivially by \(\rho_{\delta}(t)=\delta(t-\phi)\), with the
position of the collapsed probability peak \(\phi\in[0,T]\) arbitrary.
This means that, regardless of the mean delay, the whole density
collapses to a compact point discharge. Strictly speaking,
\(\rho_\delta\) is the zero-width idealization of a narrow activity
window. Its entropy is therefore understood in the coarse-grained sense,
relative to the minimal temporal resolution introduced above
(\(\Delta^{\min}=1\)), and is set to zero in this regularized
description.

It is useful here to make explicit the sense in which the linearized
point-collapse boundary is an outer estimate for the corresponding
saturating theory. Let us recall that the constant $C$ appearing in the definition of $\beta_{\textrm{MF}}=\beta C$ reads:
\[
    C=\frac{\alpha\langle W\rangle}{\gamma}.
\]
In this section \(E(\rho)\) denotes the rescaled linearized
mean-field energy \(E_{\rm lin}^{\rm tot}/(\alpha NC)\). We therefore
compare it with the analogously rescaled saturating energy
\[
    E_{\rm sat,MF}[\rho]
    =
    \frac{E_{\rm sat}^{\rm tot}[\rho]}{\alpha NC}
    =
    -\frac{1}{C}
    \int dt\;\rho(t)\tanh u_\rho(t),
\]
where
\[
    u_\rho(t)
    =
    C(\Omega*\rho)(t).
\]
The factor \(1/C\) appears because the saturating energy is rescaled by
the same factor \(\alpha NC\) that was used to define \(E_{\rm MF}\) in
Eq.~\eqref{eq:E_mf_rescaled}.
In the small-field limit, \(E_{\rm sat,MF}[\rho]\) reduces to
\(E(\rho)\).
The corresponding  rescaled saturation deficit is therefore:
\begin{eqnarray}
    g_{\rm MF}[\rho]
    &=&
    E_{\rm sat,MF}[\rho]-E(\rho)\nonumber\\
    &=&
    \frac{1}{C}
    \int dt\;\rho(t)
    \bigl[u_\rho(t)-\tanh u_\rho(t)\bigr].
\end{eqnarray}
For the uniform state,
\[
    u_{\rho_0}
    =
    \frac{C}{T},
\]
whereas for the point-collapsed state
\(\rho_\delta(t)=\delta(t-\phi)\),
\[
    u_{\rho_\delta}(\phi)
    =
    C\Omega(0).
\]
In the exponential case \(\Omega(0)=1/\mu_\Delta\). Hence, for
\(\mu_\Delta\leq T\), the point-collapsed state generates a field at
least as large as that of the uniform state. Since
\[
    \psi(u)=u-\tanh u
\]
is increasing for \(u\geq0\), it follows that
\[
    g_{\rm MF}[\rho_\delta]\geq g_{\rm MF}[\rho_0].
\]
Consequently,
\[
    E_{\rm sat,MF}[\rho_\delta]
    -
    E_{\rm sat,MF}[\rho_0]
    \geq
    E(\rho_\delta)-E(\rho_0).
\]
Thus saturation raises the relative energy of the point-collapsed state
with respect to the uniform state. Therefore, whenever the saturating
theory favours point collapse over the uniform state, the linearized
theory favours it as well. In this precise sense, the linearized point-collapse-over-uniform
boundary gives an outer estimate of the corresponding boundary for the
saturating response.

Using the linearized mean-field free energy, point collapse is expected
to occur when
\begin{align}
\label{eq:cond_md1}
\frac{1}{\mu_\Delta}
>
\frac{1}{T}
+
\frac{\log(T)}{\beta_{\rm MF}}
\end{align}
and
\begin{align}
\label{eq:cond_md2}
\frac{1}{\mu_\Delta}
>
\frac{\Omega(t^*)}{4}
+
\frac{H_\Sha}{\beta_{\rm MF}}.
\end{align}
Here \(H_{\Sha}\approx1.04\) is the maximum entropy of the three-cycle
periodic-collapsed density \eqref{eq:lam_family} with \(x=0.25\). The
transition from uniform to periodic-collapsed densities is expected to
occur when
\begin{align}
\label{eq:periodic_collapse_transition}
\frac{\Omega(t^*)}{4}
>
\frac{1}{T}
+
\frac{\log(T)-H_\Sha}{\beta_{\rm MF}},
\end{align}
and neither of the conditions \eqref{eq:cond_md1} and
\eqref{eq:cond_md2} is met (see Fig. \ref{fig:mc_mean_field}A).

\section{Monte Carlo sampling of the microscopic ensemble}
\label{sec:monte_carlo}
\subsection{Purpose: testing the mean-field phase picture}
The mean-field analysis of the previous sections predicts three broad
classes of macrostates: temporally uniform states, periodic
states selected by the delay kernel, and collapsed
discharge-like states driven by the attractive quadratic energy.
These predictions are obtained after replacing the microscopic spike
train by the population density \(\rho(t)\), averaging over the random
graph, linearizing the thresholding nonlinearity, and working with the
rescaled mean-field inverse temperature \(\beta_{\rm MF}\). The purpose
of the Monte Carlo simulations is to return to the microscopic ensemble
and test how these mean-field regimes are modified when the full
discrete raster, quenched connectivity, refractory constraints, and
nonlinear thresholding are retained. In the Monte Carlo sampler the
control parameter is the microscopic inverse temperature \(\beta\) of
the discrete ensemble, not the rescaled mean-field parameter
\(\beta_{\rm MF}\). Since the spike count, synaptic scale, and kernel
normalization are fixed in each numerical sweep, varying \(\beta\)
nevertheless varies the same effective balance between endogenous
self-consistency and entropy. Consequently, we compare the Monte Carlo and mean-field phase maps only
at the level of qualitative regimes, without attempting a pointwise
identification of \(\beta\) with \(\beta_{\rm MF}\).

\subsection{Sampling algorithm}
We now describe the Monte Carlo procedure used to sample the discrete
ensemble introduced in Sec.~\ref{sec:discrete_operator_realization}. The
goal of the simulation is not to reproduce the causal time evolution of a
neuronal network. Instead, the algorithm samples complete
spatiotemporal spike configurations from the Boltzmann distribution
\begin{equation}
    P_K(n\mid G)
    =
    Z_K^{-1}
    \exp[-\beta E_K(n)],
    \qquad
    n\in\mathcal S_K,
\label{eq:mc_boltzmann_distribution}
\end{equation}
where \(E_K(n)\) is given by Eq.~\eqref{eq:discrete_mc_energy} and
\(\mathcal S_K\) denotes the admissible set of binary trains with fixed
spike count and refractory constraints. The inverse temperature \(\beta\)
controls the relative importance of endogenous network dynamics consistency
against the entropy of admissible spike configurations. Here \(\beta\) denotes the inverse-temperature parameter of the
microscopic discrete ensemble. It should not be confused with
\(\beta_{\rm MF}\), which appears only after the annealed,
linear-response mean-field rescaling in Sec.~\ref{sec:meanfield}.

For each Monte Carlo run we first generate a directed random graph
\(G=(V,E,W,\Delta)\).  This graph is then kept fixed throughout the run
and plays the role of quenched disorder. Independent runs use
independent graph and delay realizations, unless the random seed is
explicitly fixed. Each ordered pair \(i\neq j\) is connected
independently with probability \(p_{\rm conn}\), and self-connections are
excluded. In the numerical implementation considered here all existing
synapses are assigned unit excitatory weight,
\[
    W_{ij}=1
    \quad\text{for}\quad
    (i,j)\in E,
\]
while the delays \(\Delta_{ij}\) are sampled independently for each
existing edge from a discrete-bin approximation to the causal delay
families discussed in Sec.~II\,A. In the simulations shown in
Fig.~\ref{fig:mc_mean_field}, the mean delay is fixed at
\(\mu_\Delta=5\) time bins, while the standard deviation
\(\sigma_\Delta\) is varied across the phase map. For
\(\sigma_\Delta<\mu_\Delta/2\), the numerical implementation uses a
rounded Gaussian approximation, except when the sampled delays reach the
one-bin cutoff, in which case the causal maximum-entropy sampler is used.
For \(\sigma_\Delta\ge\mu_\Delta/2\), delays are sampled from the
maximum-entropy family on \(\mathbb R_+\), with density proportional to
\(\exp(-a t-b t^2)\), where \(a\) and \(b\) are chosen to reproduce the
prescribed mean and variance. The sampled delays are then converted to
integer bins and constrained by \(\Delta_{ij}\ge1\). Thus the quenched
disorder of the simulation is carried by the adjacency matrix and the
delay matrix.

The initial microstate \(n\) is obtained by placing \(M\) spikes at
random positions in the \(N\times K\) raster and then enforcing the
absolute refractory constraint. Equivalently, the sampler is initialized
inside the canonical admissible set
\[
    \mathcal S_K
    =
    \left\{
    n\in\{0,1\}^{N\times K}:
    \sum_{i,\ell} n_{i\ell}=M,
    \quad
    n \sim\text{ refr.}
    \right\}.
\]
The fixed value of \(M\) separates the overall activity level from the
temperature parameter \(\beta\), in accordance with the canonical
convention introduced in Sec.~\ref{sec:abstract}. In terms of the mean-field notation, this fixes the analogue of the
activity scale \(\alpha\), so that changes in the Monte Carlo parameter
\(\beta\) are not confounded with changes in the total number of spikes.

At each Monte Carlo step a candidate configuration \(n'\) is generated
from the current configuration \(n\) by one of two single-attempt moves.
The first move is a spike relocation move. One occupied site
\((i,\ell)\) and one empty site \((j,m)\) are chosen uniformly. The spike
is moved from \((i,\ell)\) to \((j,m)\) only if the resulting raster
satisfies the refractory constraint; otherwise the proposal is rejected
and the state is left unchanged. The second move is a temporal walk move.
One existing spike \((i,\ell)\) is selected uniformly and proposed to move
to either \((i,\ell+1)\) or \((i,\ell-1)\), with equal probability and
with time indices understood modulo \(K\). If the target site is already
occupied or the resulting raster violates the refractory constraint, the
proposal is rejected. Both moves preserve the canonical spike count.

For every proposed configuration \(n'\), the endogenous activation field
is computed as
\[
    u'=\mathcal B_G n',
\]
using the discrete transfer operator introduced in
Sec.~\ref{sec:discrete_operator_realization} and implemented explicitly
in Appendix~\ref{app:discrete_transfer_operator}. The energy
\(E_K(n')\) is then evaluated using Eq.~\eqref{eq:discrete_mc_energy}.

The candidate configuration is accepted according to the usual
Metropolis rule
\[
    P_{\rm acc}(n\rightarrow n')
    =
    \min
    \left\{
        1,
        \exp[-\beta(E_K(n')-E_K(n))]
    \right\}.
    \label{eq:metropolis_acceptance}
\]
Each elementary move is implemented as a symmetric proposal on the
admissible state space \(\mathcal S_K\). Hence each move, and any fixed
mixture of the two moves, can be combined with the standard Metropolis
acceptance rule and satisfies detailed balance with respect to
Eq.~\eqref{eq:mc_boltzmann_distribution}. The role of combining
relocation and walk moves is not to restore proposal symmetry, but to
improve irreducibility and mixing. The temporal walk move preserves the
number of spikes assigned to each neuron and is therefore reducible on
its own, whereas the relocation move connects sectors with different
per-neuron spike counts. The resulting Markov chain is therefore used as
a sampling device for the trajectory ensemble, not as a model of
biological real-time neural dynamics.

Macroscopic observables are computed from the sampled rasters. The main
coarse-grained observable is the normalized population spike-rate profile
\[
    \rho_\ell
    =
    \frac{1}{M}
    \sum_{i=1}^{N} n_{i\ell},
    \qquad
    \sum_{\ell=0}^{K-1}\rho_\ell=1 .
\]
This is the discrete counterpart of the mean-field population-rate
density \(\rho(t)\), but it is measured only after sampling and is not a
state variable of the Markov chain. We monitor the normalized temporal
entropy
\[
    H_{\rm norm}[\rho]
    =
    -\frac{1}{\log K}
    \sum_{\ell=0}^{K-1}\rho_\ell\log\rho_\ell,
\]
and the autocorrelation of \(\rho_\ell\). Entropy provides
 continuous diagnostics of the temporal localization, while
nonzero autocorrelation peaks distinguish repeated activity windows from
a single temporally localized discharge.

The Monte Carlo simulations thus probe the competition between the
energy term, which favors spike configurations self-consistent with the
delayed transfer operator \(\mathcal B_G\), and the entropy of admissible
trains, which favors temporally disordered states. In this
sense the simulations are a direct numerical realization of the
statistical-mechanical ensemble over spike trains, while remaining a
deliberately reduced surrogate of explicit membrane-based neuronal
dynamics.

\subsection{Results of the simulations and relation to the mean-field theory}
\label{sec:mc_results_mean_field}

The linearized mean-field free energy
\(\mathcal F_{\rm lin}[\rho]\) is obtained by projecting the microscopic
ensemble onto the population-rate density \(\rho(t)\), replacing the
quenched random graph by its annealed delay-response kernel, and
rescaling the microscopic inverse temperature into
\(\beta_{\rm MF}\). The Monte Carlo simulations instead sample binary
trains \(n_{i\ell}\) in finite networks at fixed microscopic
\(\beta\). Each individual Markov chain is run at a fixed quenched realization of
the graph \(G\). In the phase maps shown below, different points of the
parameter grid are obtained from independent runs in which the graph and
delay assignment are sampled anew. Thus the maps should not be read as
disorder-averaged observables at each parameter point, but as a
single-realization finite-size phase portrait of the sampled microscopic
ensemble. The observables shown in the phase map are computed from the final
sampled state of each chain, rather than from an explicit average over
many independent chains at the same parameter values.

The population density
\[
    \rho_\ell
    =
    M^{-1}\sum_i n_{i\ell}
\]
is measured only after sampling and is therefore an observable of the
microscopic ensemble, not a state variable of the simulation. This
distinction is important: the mean-field theory studies the free energy
of population-rate macrostates, whereas the Monte Carlo procedure samples
finite spike trains from the underlying microstate ensemble.

\begin{figure*}
    \centering
    \includegraphics[
        width=\textwidth, trim=0 400 0 0,
    clip]{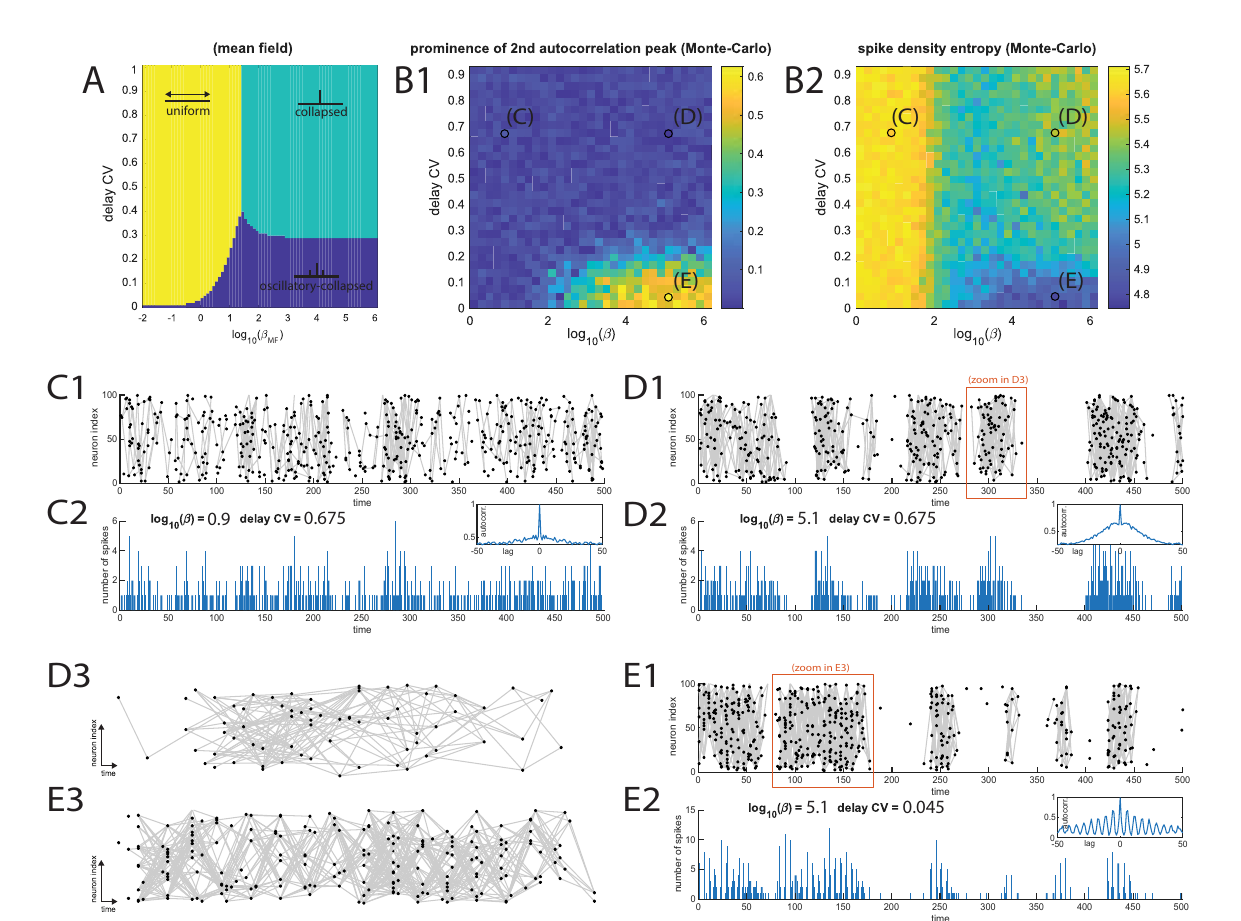}
    \caption{ Comparison of mean-field predictions and Monte Carlo results. (A) Three-regime mean-field phase diagram obtained by combining the
uniform, point-collapsed, and oscillatory-collapsed criteria discussed in
Sec.~\ref{sec:phases}. (B) maps of the second autocorrelation-peak prominence (B1) and
spike-rate entropy (B2), computed from the final Monte Carlo state at
each parameter point. The qualitative characteristics of phases and their relative location in parameter space are maintained. (C) final Monte Carlo state for uniform phase, spike train with links representing energy-contributing interactions (C1) and the corresponding spike rate distribution (C2), inset showing associated autocorrelation function. (D1-2) point-collapsed phase, sub-panels as above; note the 'glassy' disordered internal structure of spike interactions qualitatively similar to Izhikevich's neuronal groups (zoom  panel D3). (E) oscillatory-collapsed phase, sub-panels as above, note the temporally ordered internal structure of spike interactions (zoom panel E3).}
\label{fig:mc_mean_field}
\end{figure*}

Figure~\ref{fig:mc_mean_field} compares the resulting Monte Carlo phase
map with the mean-field prediction. Figure~\ref{fig:mc_mean_field}A
shows the phase diagram obtained from the coarse-grained free-energy
analysis, while Fig.~\ref{fig:mc_mean_field}B shows the corresponding
classification obtained from the sampled microscopic rasters. The comparison is qualitative rather than pointwise: the Monte Carlo
sampler does not solve the mean-field equations, and the bare Monte
Carlo parameter \(\beta\) is not identical to the rescaled mean-field
parameter \(\beta_{\rm MF}\). Agreement between the two maps therefore
indicates that the dominant organization of the sampled rasters is
controlled by the same competition between delayed self-consistency and
temporal entropy that appears in \(\mathcal F_{\rm lin}[\rho]\).

The main mean-field regimes are visible in the sampled ensemble. At small microscopic \(\beta\), corresponding to a high-temperature or
weak-structure regime of the sampled ensemble, spike events are
distributed throughout the observation window and the population-rate
profile remains broadly dispersed (Fig.~\ref{fig:mc_mean_field}C). This corresponds to the uniform
macrostate of the mean-field theory. At larger \(\beta\), or for delay
distributions that favour strong recurrent self-consistency, the sampled
rasters become temporally localized (Fig.~\ref{fig:mc_mean_field}D-E). This realizes the collapse tendency
predicted by the linearized quadratic mean-field energy. The distinction
between oscillatory and non-oscillatory localized regimes is therefore
not simply a matter of stronger or weaker localization. Rather, it is a
distinction between temporally ordered localization, where activity forms
repeated windows separated by a delay-selected period, and temporally
disordered localization, where activity remains localized but lacks a
regular recurrence structure (compare Fig.~\ref{fig:mc_mean_field} D3 vs E3). The former corresponds to the
oscillatory-collapsed regime of the mean-field picture, while the latter
is closer to a glassy non-oscillatory collapsed regime.

The Monte Carlo phase map also shows systematic deviations from the
idealized mean-field prediction. The boundaries between regimes are broadened rather than sharp, as
expected for finite systems and finite sampling times. Additional
irregularity is introduced by the use of a single quenched
graph-and-delay realization at each parameter point.  More
importantly, complete temporal collapse is suppressed in the microscopic
ensemble. Several mechanisms contribute to this effect. First, refractory exclusion is represented in the mean-field picture only
through a coarse temporal cutoff, whereas in the microscopic Monte Carlo
ensemble it is imposed as a neuron-resolved hard constraint. This limits
how densely spikes from the same neuron can be packed in time and
suppresses the idealized complete temporal collapse allowed by the
one-density mean-field description. Second, the nonlinear thresholding function introduces
saturation: once a spike is sufficiently supported by converging
presynaptic activity, additional presynaptic input no longer lowers the
energy proportionally. Third, quenched graph disorder and finite
connectivity preserve neuron-specific network structure that is absent from a
description based only on \(\rho(t)\). These effects redistribute
activity over several time bins, over multiple cycles of an oscillation,
or over neuron-specific activity groups. This is consistent with the interpretation of the linearized collapse
boundaries in Sec.~\ref{sec:phases} as outer estimates: the linearized
mean-field theory over-rewards large endogenous fields, while the
microscopic simulations include saturation and hard admissibility
constraints that reduce the tendency toward complete collapse.

Thus the Monte Carlo simulations support the mean-field classification
at the level of broad macroscopic regimes, but also identify the
limitations of the annealed, linearized, one-density approximation. The
mean-field theory captures the energetic tendency toward uniform,
periodic, or collapsed temporal organization, whereas the microscopic
ensemble reveals how this tendency is reshaped by refractory constraints,
threshold saturation, finite-size fluctuations, and graph-specific
structure. The formation of
structured activity groups in the sampled trains is therefore not a
failure of the statistical-mechanical framework, but rather a sign that
the relevant macrostates cannot always be parametrized by the
population-rate density alone.

\section{Inferring connectivity}
\label{sec:conn}
One advantage of the proposed framework is that it can be inverted and
used to infer connectivity parameters from an empirical spike train. In
this section the spike train \(s\) is fixed, and the ensemble is defined
over a family \(\mathfrak A\) of admissible intrinsic-response operators
\(\mathcal A\), or equivalently over graph/connectivity parameters. The
constraints defining \(\mathfrak A\) represent prior knowledge about the
system, such as the connection probability \(p_{\rm conn}\), the delay
distribution \(P_\Delta\), and the synaptic-weight distribution \(P_W\).

For a candidate operator \(\mathcal A\), let
\[
    f_i(t)
    =
    \mathcal T[(\mathcal A s)_i(t)]
\]
be the thresholded endogenous activation field generated by the observed
spike train. We define the compatibility energy as
\begin{equation}
    E_{\rm inf}(\mathcal A;s)
    =
    -
    \sum_{i=1}^{N}
    \int_0^T dt\,
    s_i(t)\,
    \mathcal T[(\mathcal A s)_i(t)] .
\label{eq:Einf}
\end{equation}
Equivalently, since \(s_i(t)\) is a spike train, this can be written as
\[
    E_{\rm inf}(\mathcal A;s)
    =
    -
    \sum_{i=1}^{N}
    \sum_{t\in {\rm sp}(i)}
    \mathcal T[(\mathcal A s)_i(t)] .
\]
Thus a candidate connectivity model receives lower energy when the
observed spikes occur at times at which the same model generates strong
endogenous activation.

We then define the inverse ensemble
\begin{equation}
    P(\mathcal A\mid s)
    =
    Z_s^{-1}
    \exp[-\beta_{\rm inf}E_{\rm inf}(\mathcal A;s)],
\end{equation}
with
\[
    Z_s
    =
    \sum_{\mathcal A\in\mathfrak A}
    \exp[-\beta_{\rm inf}E_{\rm inf}(\mathcal A;s)] .
\]
The parameter \(\beta_{\rm inf}\) is an inference precision parameter:
large \(\beta_{\rm inf}\) favours operators that make the observed spike
train strongly self-consistent, whereas small \(\beta_{\rm inf}\) leaves
the posterior dominated by the prior constraints. It is conceptually
related to the microscopic inverse temperature \(\beta\), but it is not
identified with it unless an additional calibration is imposed. Stability
of the inferred structure under changes of \(\beta_{\rm inf}\) may be
used as a robustness diagnostic.

Let us consider the simplest situation, in which the observed spike train
\(s\) contains spikes from only two neurons, \(a\) and \(b\), and we want
to infer whether there is an excitatory connection from \(a\) to \(b\).
Since in this setup excitatory postsynaptic potentials in \(b\) arising
from the putative connection are generally not expected to reach the
spike-generation threshold on their own, we may again use the
weak-coupling assumption~\eqref{eq:VtanhV}. Let us further assume that we have prior knowledge about the weight $W$ and delay $\Delta$ of the connection and that neither $b\rightarrow a$ nor self-connections are present. Hence, only two candidate models are considered (connection present "$+$" and connection absent "$-$"), and their energies are:
\begin{align}
\label{eq:Eplus}
E_{+}=-W\,\sum\limits_{\begin{smallmatrix}t\,\in sp(a)\\t'\in sp(b) \end{smallmatrix}} \eta(t'-t-\Delta)
\end{align}
\begin{align}
E_{-}=0
\end{align}
where $\eta$ is the IAF voltage response kernel \eqref{eq:eta_IAF}. The partition function is
\[
    Z_s=\exp(-\beta_{\rm inf}E_+)+\exp(-\beta_{\rm inf}E_-)
       =\exp(-\beta_{\rm inf}E_+)+1,
\]
and therefore
\[
    P(a\to b\mid s)
    =
    \frac{\exp(-\beta_{\rm inf}E_+)}
    {\exp(-\beta_{\rm inf}E_+)+1}.
\]
The choice of \(\beta_{\rm inf}\) controls the sharpness of the inference. In the limit \(\beta_{\rm inf}\to0\), the observation does not
distinguish the two candidate models and the posterior probability is
\(1/2\). In the limit \(\beta_{\rm inf}\to\infty\), the lower-energy
candidate dominates; for \(E_+<0\), this is the model containing the
connection. Note that the expression \eqref{eq:Eplus} is similar to the total amount of long-term potentiation induced at the putative synapse in the spike-timing dependent plasticity (STDP) framework \cite{Song2000}. 

For the second example, let us switch to the population-rate
representation of Sec.~\ref{sec:meanfield}, using the trivial response
kernel in Eq.~\eqref{eq:trivkernel}. We ask how one could infer a
normalized causal delay kernel \(\Omega\) from an empirical
population-rate density \(\rho\). Here \(\Omega\) is the inferred object,
while \(\rho\) is fixed.

Up to an overall positive scale absorbed into \(\beta_{\rm inf}\), the
compatibility energy is
\begin{equation}
    E_\Omega[\Omega;\rho]
    =
    -\int dt\;\Omega(t)R_{\rho\rho}(t),
\end{equation}
where
\begin{equation}
    R_{\rho\rho}(t')
    =
    \int dt\;\rho(t)\rho(t-t')
\end{equation}
is the autocorrelation of \(\rho\). The entropy associated with kernels
having the same coarse-grained delay distribution is written as
\begin{equation}
    H[\Omega]
    =
    -\int dt\;\Omega(t)\log\Omega(t)+H_0,
\end{equation}
where \(H_0\) collects connectivity and weight-distribution factors
independent of the shape of \(\Omega\). The corresponding inverse
free-energy functional is
\begin{equation}
    F_\Omega[\Omega;\rho]
    =
    E_\Omega[\Omega;\rho]
    -
    \frac{H[\Omega]}{\beta_{\rm inf}}
\end{equation}
or, up to the irrelevant constant \(-H_0/\beta_{\rm inf}\),
\begin{equation}
\label{eq:Ffunc}
F_\Omega[\Omega;\rho]
    =
    -\int dt\;\Omega(t)R_{\rho\rho}(t)
    +
    \frac{1}{\beta_{\rm inf}}
    \int dt\;\Omega(t)\log\Omega(t).
\end{equation}
Since $\beta_{\mathrm{inf}}$ is a free precision parameter, the overall positive scale absorbed here (the fixed factors $N^{2}$, $p_{\mathrm{conn}}$, and $\langle W\rangle$) merely rescales $\beta_{\mathrm{inf}}$ relative to the two-neuron example and does not affect the result.
 
 The functional is minimized over normalized causal kernels satisfying
\begin{equation}
    \int dt\,\Omega(t)=1,
    \qquad
    \Omega(t\le0)=0,
\end{equation}
and, if imposed, moment constraints such as
\begin{equation}
    \int dt\,t\,\Omega(t)=\mu_\Delta,
    \qquad
    \int dt\,t^2\Omega(t)-\mu_\Delta^2=\sigma_\Delta^2.
\end{equation}
With only normalization imposed, the stationary solution is
\begin{equation}
    \Omega^*(t)
    \propto
    \exp\!\left[
        \beta_{\rm inf}R_{\rho\rho}(t)
    \right].
\end{equation}
With moment constraints included, this generalizes to
\begin{equation}
    \Omega^*(t)
    \propto
    \exp\!\left[
        \beta_{\rm inf}R_{\rho\rho}(t)
        -\lambda_1 t-\lambda_2 t^2
    \right],
\end{equation}
with the appropriate multipliers omitted when the corresponding moment
constraint is not imposed.

Thus the inferred kernel places probability mass at delays where the
empirical population-rate density has large autocorrelation, modulated by
the maximum-entropy constraints on the delay distribution. In the trivial case of a uniform empirical density,
\(R_{\rho\rho}\) is constant. The autocorrelation term then carries no
delay-selective information, and the solution reduces to the
maximum-entropy causal delay distribution determined solely by the
imposed normalization and moment constraints, in the same sense as the
delay families introduced in Sec.~\ref{sec:ensemble} A and used in
Eqs.~\eqref{eq:Gaussian}--\eqref{eq:exponential_dist}.

\section{Discussion and conclusions}
\label{sec:disc}
We have introduced a statistical-mechanical framework in which complete spike
trains, rather than underlying neuronal state variables, constitute the microscopic
degrees of freedom of the system. This perspective differs fundamentally from most
existing approaches in theoretical neuroscience, which begin with dynamical
equations for membrane potentials, conductances, adaptation variables, or
population rates and only subsequently derive spike statistics \cite{Montbri015, diVolo19, Chen22, Tahvili24, Tuckwell1988, Burkitt2006a, Burkitt2006b, Gerstner2014}. Here,
the spike train itself is the primary object of the theory. To our knowledge, this
trajectory-ensemble formulation represents a novel statistical-physics description of
neuronal activity, allowing neural systems to be analyzed directly in terms of
experimentally observable quantities.
An important advantage of this construction is its direct applicability to experimental data measurable using modern setups. Internal neuronal state variables are generally inaccessible in vivo and often
only partially observable even in detailed electrophysiological preparations. In
contrast, spike trains can be measured directly using multielectrode arrays \cite{Giraud25},  calcium-imaging techniques, and modern optogenetic approaches \cite{Andrews24}. Because the
formalism is defined on spike histories rather than on latent dynamical states, it
naturally lends itself to data-driven analysis and inference without requiring
reconstruction of unobservable membrane or network variables \cite{Schneidman2006, Tkacik2013, Cocco2009, Roudi2009}. Even though
the energy functional depends on an assumed network model, the neuronal heat
bath accommodates, to an extent, uncertainties about its parameters along with
external interactions.
Despite its abstraction, the framework captures several biologically relevant collective
phenomena. In particular, it predicts that, in purely excitatory systems, the
competition between energetic self-consistency and configurational entropy gives rise
to temporally localized discharge-like states and population bursts. Such bursting
phenomena are ubiquitous in neural systems and are implicated both in physiological
processes, including development and memory consolidation as well as
pathological dynamics such as epileptic interictal discharges and seizures \cite{BenAri14, Buzsaki06, Jiruska13, Luscher12}.
The emergence of these states as distinct phases of the trajectory ensemble
suggests that bursting may be understood as a generic collective phenomenon
arising from the statistical organization of spike trains.

The present analysis is restricted to homogeneous excitatory delayed
networks. Within this scope, the framework captures both delay-selected
oscillatory organization and temporally localized discharge-like states
as competing minima of the spike-train free-energy landscape. This places
population bursting, rhythmic temporal order, and non-oscillatory
localized activity within a common statistical-mechanical description of
spike-train space.

A notable feature of the Monte Carlo ensemble is the emergence of structured
activity groups rather than complete temporal collapse. These activity clusters
resemble neuronal assemblies and polychronous groups described previously in
large-scale spiking network simulations, particularly those reported by Izhikevich \cite{Izhikevich10, Izhikevich06}. Their appearance indicates that finite-size effects, refractory constraints, threshold
nonlinearities, and quenched network structure can stabilize rich mesoscopic
organization beyond the simplest mean-field picture. The reproduction of such
grouped activity patterns suggests that the proposed ensemble retains important
features of biologically realistic spiking networks despite its reduced description.
From a broader neuroscience perspective, the observed transitions between
asynchronous, oscillatory, and burst-like phases may be especially significant
because neuronal systems are widely believed to operate near dynamical
bifurcations or critical phase transitions \cite{Beggs03, Shew13, Kinouchi06, Hesse14, Munoz18, Chialvo10}. A substantial body of theoretical and
experimental work has suggested that proximity to critical points can optimize
information transmission, dynamic range, sensitivity to stimuli, computational
capacity, and flexibility of network responses \cite{Beggs03, Shew13, Kinouchi06, Hesse14, Munoz18, Chialvo10, Mora11}. Within the present framework,
these notions acquire a natural statistical-mechanical interpretation: different modes
of neural computation correspond to distinct regions of the free-energy landscape,
while changes in information-processing capabilities may be associated with
proximity to phase boundaries, allowing for metastability.
Perhaps the most striking conclusion emerging from the analysis is the central role
played by synaptic delays. In the mean-field reduction, the effective delay kernel
appears as an irreducible quantity governing the free-energy landscape. The
locations and existence of phase transitions are determined primarily by the shape of
this kernel rather than by microscopic details of individual neurons. Narrow delay
distributions favor oscillatory states, whereas broader distributions promote
temporally localized discharge-like states. While synaptic delays are known to
influence synchronization, oscillations, and polychronization \cite{Izhikevich06, Roxin05, Vicente08, Deco12}, our results
suggest that they may constitute one of the fundamental organizing variables of
collective neural activity and therefore deserve greater attention in both theoretical
and experimental studies.

Several limitations remain. The current analysis relies on simplifying assumptions,
including weak-coupling approximations, homogeneous populations, and idealized
delay distributions. Extending the framework to heterogeneous cell classes including inhibitory interneurons, multi-class
synaptic interactions, structured connectivity, plastic synapses, and
experimentally inferred interaction kernels represents an important
direction for future work.

To conclude, statistical mechanics defined directly on spike-trains is sufficient to generate a rich
phenomenology encompassing asynchronous activity, oscillations, neuronal
assemblies, and bursting states. More generally, the framework establishes a bridge
between modern statistical physics and experimentally accessible neural activity,
providing a route toward ensemble-based inference, model comparison, and the
characterization of phase transitions in neuronal systems.

\section*{Acknowledgements}
Computations were carried out using the computers of the Center of Informatics Tricity Academic Supercomputer $\&$ Network (TASK, grant number pt01269)
\section*{Data Availability Statement}
The Monte Carlo simulation code is publicly available in Ref.~\cite{repo_mc}. The mean-field results shown in Fig.~1 can be reproduced directly from the equations presented in this article; no separate implementation is provided. No experimental or personal data were used in this study.

\section*{AI Disclosure}
During manuscript preparation, the authors used ChatGPT to assist with
language editing, structural revision, and consistency checks. All
scientific content, calculations, interpretations, and final text were
reviewed and approved by the authors.

\bibliography{statMechNeuro}

\appendix

\section{Fourier representation}
\label{app:fourier}

We present a Fourier representation of the spike-train ensemble. The
purpose of this appendix is to diagonalize the linear intrinsic-response
operator in the temporal variable. The thresholding nonlinearity remains
pointwise in time and therefore couples Fourier modes.

For any sufficiently regular test function \(f_i(t)\), the action of a
spike train on \(f\) is simply evaluation at the spike times:
\begin{equation}
    \sum_{i=1}^{N}
    \int_0^T dt\,
    s_i(t)f_i(t)
    =
    \sum_{i=1}^{N}
    \sum_{t\in{\rm sp}(i)}
    f_i(t).
\label{eq:spike_train_evaluation}
\end{equation}
With this convention the energy functional can be written as
\begin{equation}
    E(s)
    =
    -
    \sum_{i=1}^{N}
    \int_0^T dt\,
    s_i(t)\,
    \mathcal T[(\mathcal A s)_i(t)] .
\end{equation}
The intrinsic response operator is given by
\[
    (\mathcal A s)_j(t)
    =
    \sum_{i\in{\rm pre}(j)}
    W_{ij}(s_i*\eta)(t-\Delta_{ij}) .
\]

Assuming periodic boundary conditions on the observation window
\([0,T]\), we introduce the Fourier coefficients
\[
    \psi_i(q)
    =
    \int_0^T dt\,
    s_i(t)
    \exp\!\left(
        -\frac{2\pi {\rm i}q t}{T}
    \right),
    \qquad q\in\mathbb Z .
\]
For a spike train this gives
\[
    \psi_i(q)
    =
    \sum_{t\in sp(i)}
    \exp\!\left(
        -\frac{2\pi {\rm i}q t}{T}
    \right).
\]

The linear response operator is diagonal in the Fourier index \(q\):
\[
    \widehat{(\mathcal A s)}_j(q)
    =
    \sum_i
    \widehat{\mathcal A}_{ji}(q)\psi_i(q),
\]
with
\[
    \widehat{\mathcal A}_{ji}(q)
    =
    W_{ij}\,
    \widehat\eta(q)\,
    \exp\!\left(
        -\frac{2\pi {\rm i}q\Delta_{ij}}{T}
    \right).
\]
For the IAF response kernel
\[
    \eta(t)=\kappa\Theta(t)e^{-t/\tau_m},
\]
the corresponding infinite-line Fourier transform is
\[
    \widehat\eta(\omega)
    =
    \frac{\kappa}{1/\tau_m+2\pi {\rm i}\omega}.
\]
On a finite periodic window one should instead use the corresponding
Fourier-series coefficients of the periodically continued kernel.

The thresholding map is pointwise in time,
\[
    f(t)\mapsto \mathcal T[f(t)],
\]
and therefore becomes a nonlinear, mode-coupling map in Fourier space:
\[
    \widetilde{\mathcal T}[\psi]
    =
    F\left[
        \mathcal T(F^{-1}\psi)
    \right].
\]
Consequently the energy in Fourier variables is most safely written by
transforming back to the time domain before applying the thresholding
map:
\begin{equation}
    E(\psi)
    =
    -
    \sum_{j=1}^{N}
    \int_0^T dt\,
    (\mathcal F^{-1}\psi)_j(t)\,
    \mathcal T\!\left[
        (\mathcal F^{-1}\widehat{\mathcal A}\psi)_j(t)
    \right].
\label{eq:fourier_energy_explicit}
\end{equation}

The Boltzmann weight is therefore
\[
    P(\psi)
    =
    Z^{-1}
    \exp[-\beta E(\psi)].
\]

In the weak-coupling limit, where the thresholding nonlinearity is
linearized, the energy becomes quadratic in the Fourier amplitudes and
the diagonal structure of \(\widehat{\mathcal A}_{ji}(q)\) can be used
directly. Outside this limit, the Fourier representation remains useful
for the linear propagation step, but the thresholding operation couples
different Fourier modes.

\section{Discrete transfer operator used in Monte Carlo}
\label{app:discrete_transfer_operator}

In the discrete Monte Carlo realization of
Sec.~\ref{sec:discrete_operator_realization}, the transfer operator
\(\mathcal B_G\) is implemented in the temporal Fourier representation.
We first take the discrete Fourier transform of the raster along the
time-bin index,
\begin{equation}
\label{eq:dft_def}
    \widehat n_i(q)
    =
    \sum_{\ell=0}^{K-1}
    n_{i\ell}
    \exp\!\left(
        -\frac{2\pi {\rm i}q\ell}{K}
    \right),
    \qquad
    q=0,\ldots,K-1.
\end{equation}
For each Fourier mode \(q\), delayed propagation through the graph is
encoded by an \(N\times N\) matrix \(\widehat B(q)\). We define the
Fourier components of the endogenous activation field by
\[
    \widehat u_j(q)
    =
    \sum_{i=1}^{N}
    \widehat B_{ji}(q)\,\widehat n_i(q),
\]
with
\begin{equation}
    \widehat B_{ji}(q)
    =
    \widehat h(q)\,
    A_{ij}\,
    \exp\!\left(
        -\frac{2\pi {\rm i}q\,\Delta_{ij}}{K}
    \right).
\label{eq:discrete_transfer_operator}
\end{equation}
With the Fourier convention
\eqref{eq:dft_def}, the negative sign in
Eq.~\eqref{eq:discrete_transfer_operator} implements the causal delay
\(s_i(t-\Delta_{ij})\).

Here \(h\) is an effective temporal response kernel and \(\widehat h\) is
its discrete Fourier transform. In the simulations below we use
\begin{equation}
    h_\ell
    =
    \exp\!\left(-\frac{\ell}{5}\right)
    -
    \frac{1}{2}
    \exp\!\left(-\frac{K-1-\ell}{5}\right),\,\,
    \ell=0,\ldots,K-1,
\label{eq:mc_temporal_kernel}
\end{equation}
which matches the kernel used in the numerical implementation. The
second term compensates the periodic boundary convention implicit in the
Fourier representation and suppresses a purely one-sided wrap-around
response. With the Fourier convention of Eq.~\eqref{eq:dft_def}, the discrete
Fourier transform of this kernel is
\[
    \widehat h(q)
    =
    \sum_{\ell=0}^{K-1}
    h_\ell
    \exp\!\left(
        -\frac{2\pi {\rm i}q\ell}{K}
    \right),
    \qquad q=0,\ldots,K-1.
\]
Equivalently, writing
\[
    r=e^{-1/5},
    \qquad
    z_q=
    \exp\!\left(
        -\frac{2\pi{\rm i}q}{K}
    \right),
\]
one obtains
\begin{equation}
    \widehat h(q)
    =
    (1-r^K)
    \left[
        \frac{1}{1-r z_q}
        -
        \frac{1}{2}
        \frac{z_q^{-1}}{1-r z_q^{-1}}
    \right].
\label{eq:mc_temporal_kernel_fourier}
\end{equation}
In the MATLAB implementation this quantity is stored as
\(\texttt{kern(q+1)}\), due to one-based indexing.

The time-domain endogenous activation field is then obtained by the
inverse discrete Fourier transform,
\begin{equation}
    u_{j\ell}
    =
    (\mathcal B_G n)_{j\ell}
    =
    \frac{1}{K}
    \sum_{q=0}^{K-1}
    \widehat u_j(q)
    \exp\!\left(
        \frac{2\pi {\rm i}q\ell}{K}
    \right).
\label{eq:inverse_fourier_activation}
\end{equation}
Equivalently,
\[
    \mathcal B_G n
    =
    \operatorname{Re}\,
    \mathcal F_t^{-1}
    \left[
        \widehat B(q)\,\widehat n(q)
    \right],
\]
where the inverse transform acts only on the temporal Fourier index. The
operator \(\operatorname{Re}\) reflects the numerical implementation, in
which the real part of the inverse transform is retained as the
activation field entering the energy.
\end{document}